\documentclass[fleqn,usenatbib]{mnras}

\usepackage{newtxtext,newtxmath}
\usepackage{graphicx}
\usepackage{caption}
\usepackage{subcaption}
\graphicspath{ {./Figures/} }

\usepackage[T1]{fontenc}

\DeclareRobustCommand{\VAN}[3]{#2}
\let\VANthebibliography\thebibliography
\def\thebibliography{\DeclareRobustCommand{\VAN}[3]{##3}\VANthebibliography}

\usepackage{graphicx}	
\usepackage{amsmath}	

\title[Joint NuSTAR/STIX HXR Microflares]{First joint X-ray solar microflare observations with NuSTAR and Solar Orbiter/STIX}

\author[N. Bajnoková et al.]{
Natália Bajnoková,$^{1}$\thanks{E-mail: n.bajnokova.1@research.gla.ac.uk}
Iain G. Hannah,$^{1}$
Kristopher Cooper,$^{2}$
Säm Krucker,$^{3, 4}$ 
Brian W. Grefenstette,$^{5}$
\newauthor{
David M. Smith,$^{6}$
Natasha L. S. Jeffrey$^{7}$,
Jessie Duncan$^{8}$
}
\\
$^{1}$School of Physics \& Astronomy, University of Glasgow, University Avenue, Glasgow G12 8QQ, UK\\
$^{2}$School of Physics \& Astronomy, University of Minnesota Twin Cities, Minneapolis, MN 55455, USA\\
$^{3}$School of Engineering, University of Applied Sciences and Arts Northwestern Switzerland, CH-5210 Windisch, Switzerland\\
$^{4}$Space Sciences Laboratory, University of California, Berkeley, CA 94720, USA\\
$^{5}$Cahill Center for Astrophysics, California Institute of Technology, 1216 East California Boulevard, Pasadena, CA 91125, USA\\
$^{6}$Santa Cruz Institute of Particle Physics and Department of Physics, University of California, Santa Cruz, CA 95064, USA\\
$^{7}$Department of Mathematics, Physics \& Electrical Engineering, Northumbria University, Newcastle upon Tyne NE1 8ST, UK\\
$^{8}$NASA Marshall Space Flight Center, ST13, Huntsville, AL 35812, USA\\
}

\date{Accepted XXX. Received YYY; in original form ZZZ}

\pubyear{2023}

\begin{document}
\label{firstpage}
\pagerange{\pageref{firstpage}--\pageref{lastpage}}
\maketitle

\begin{abstract}

We present the first joint spectral and imaging analysis of hard X-ray (HXR) emission from 3 microflares observed by the Nuclear Spectroscopic Telescope ARray (NuSTAR) and Solar Orbiter/Spectrometer/Telescope for Imaging X-rays (STIX). We studied 5 joint spectra from GOES A7, B1 and B6 class microflares from active region AR12765 on 2020 June 6 and 7. As these events are very bright for NuSTAR, resulting in extremely low (<1\%) livetime,  we introduce a pile-up correction method. All five joint spectra were fitted with an isothermal model finding temperatures in the 9--11 MK range. Furthermore, three joint spectra required an additional non-thermal thick-target model finding non-thermal powers of $10^{25}$--$10^{26}$~erg~s$^{-1}$. All the fit parameters were within the ranges expected for HXR microflares. The fit results give a relative scaling of STIX and NuSTAR mostly between 6-28\% (one outlier at 52\%) suggesting each instrument are well calibrated. In addition to spectral analysis, we performed joint HXR imaging of the June 6 and one of the June 7 microflares. In NuSTAR's field of view (FOV), we observed two separate non-thermal sources  connected by an elongated thermal source during the June 6 microflares. In STIX's FOV (44\textdegree{} W with respect to NuSTAR), we imaged thermal emission from the hot flare loops which when reprojected to an Earth viewpoint matches the thermal sources seen with NuSTAR and in the hotter EUV channels with the Solar Dynamic Observatory's Atmospheric Imaging Assembly.

\end{abstract}




\section{Introduction}

The impulsive energy release in solar flares is thought to be a result of magnetic reconnection in the solar corona \citep{Fletcher_2011}, resulting in particle acceleration and plasma heating \citep[e.g.,][]{Benz_2017}, thus producing X-ray thermal and non-thermal bremsstrahlung emission. Microflares, defined as GOES B class and smaller active region (AR) flares several orders of magnitudes fainter than the big flares (but not necessarily spatially small), are thought to follow the same underlying processes \citep{Hannah_2011}.

The first comprehensive statistical studies of over 24,000 X-ray microflares were conducted using observations from the Reuven Ramaty High Energy Solar Spectroscopic Imager (RHESSI) \citep[e.g.][]{Christe_2008, Hannah_2011}. RHESSI was an X-ray imaging spectrometer that observed the Sun from 3 keV to 17 MeV and operated from 2002 to 2018 \citep{rhessi_paper}. RHESSI microflares revealed that, like larger flares, they occur in active regions. X-ray imaging often showed elongated loop structures with higher energy emission emanating from the flare footpoints. The microflares showed temperatures within $10.7 \leq \textrm{T} \leq 15.5$ MK and emission measures of $10^{45} \leq \textrm{EM} \leq 10^{47} \textrm{cm}^{-3}$. Typical non-thermal powers were found to be within $10^{25}$--$10^{28}$ erg s$^{-1}$ \citep{Hannah_2008}. Other hard X-ray (HXR) instruments that observed microflares in recent years are the Nuclear Spectroscopic Telescope ARray \citep[NuSTAR;][]{Harrison_2013}{} and Solar Orbiter/ Spectrometer/Telescope for Imaging X-rays \citep[STIX;][]{Krucker_2020}{}.

NuSTAR is a NASA Small Explorer mission with direct focusing optics that is capable of observing HXR sources in 3--79 keV range; however, during solar observations this energy limit can be extended down to 2.5 keV \citep{Glesener_2020}. NuSTAR mirrors focus photons onto two focal plane modules FPMA and FPMB. Each focal plane consists of 4 detector chips and has a field of view (FOV) of 12' $\times$ 12' \citep{Harrison_2013}. NuSTAR was primarily designed to observe astrophysical sources, however since 2014 it has also been observing the Sun several times per year. NuSTAR has provided a unique opportunity to observe wide range of solar activity from faint quiet sun features \citep{Paterson_2023} up to larger GOES A and B class microflares \citep[e.g.,][]{Iain_2016, Hannah_2019, Glesener_2020, Cooper_2020}. As NuSTAR is optimized for non-solar observations, its low detector throughput, and hence low livetime during solar observations, can result in spectral artefacts such as pile-up (distortion in measured photon energies due to additional photons arriving during the processing time of a prior photon detection) \citep{Grefenstette_2016}.

STIX is a HXR instrument onboard Solar Orbiter. It uses an indirect imaging technique \citep{Massa_2023} and covers an energy range of 4--150 keV. Due to its elliptical orbit around the Sun, it provides a unique opportunity to detect flares as close as 0.3~au and from a different perspective than telescopes at Earth. STIX continuously observes the Sun and its FOV covers the entire solar disk. Since its launch in 2020, it has observed over 40,000 flares, including both GOES A and B class microflares \citep[][]{Battaglia_2021, Saqri_2022}.

There are several advantages for conducting joint HXR microflare observations with NuSTAR and STIX. Combined observations of the same event observed by the two telescopes give us better constraint on the physical parameters of the heated plasma and accelerated particles while also allowing us to check the calibration of each instruments. Due to the difference in STIX and NuSTAR's orbit, we have the possibility to observe the same event from different angles; for example, observing both the faint coronal source and bright footpoints which is not yet possible with a single HXR instrument \citep{Krucker_2019}. STIX does not suffer significantly from the effects of pile-up like NuSTAR which can obscure the non-thermal component in an existing spectra. On the other hand, STIX has lower sensitivity than NuSTAR. Furthermore, for lower GOES B and A class flares, the STIX spectrum is dominated by the emission from the onboard calibration source \citep{Battaglia_2021}. Thus, with the joint observations, we can get a better constraint on the higher energy range with STIX while still constraining the lower energy, thermal part of the spectrum with NuSTAR. Lastly, since the observed on-disk events are at the limits of each instrument, the joint analysis allows us to get a reliable fit which we might not be able to achieve from analysis of the individual spectra.

This paper provides the first spectral analysis of 5 joint spectra from 3 microflares simultaneously observed by NuSTAR and STIX on 2020 June 6 and 7. In section \ref{sec:overview}, we provide an overview of the instrumentation and the microflare observation campaign and discuss it the context of EUV emission observed by the Solar Dynamics Observatory’s Atmospheric Imaging Assembly \citep[SDO/AIA;][]{Lemen_2012}{}{}. In section \ref{sec:pile-up}, we introduce a new method for the first-order NuSTAR pile-up correction that is used for all NuSTAR spectral analysis throughout this study. Sections \ref{sec:joint_fitting_06} and \ref{sec:joint_imaging} include detailed joint spectral and imaging analysis of the June 6 microflare. In addition, we discuss the necessary corrections required for NuSTAR spectral fitting and the process of joint spectral fitting. We provide joint spectral and imaging 
analysis of the June 7 microflares in Section \ref{sec:joint_fitting_07}.

\section{Overview of Observations}
\label{sec:overview}

\begin{figure}
\centering
\includegraphics[height=4.5cm]{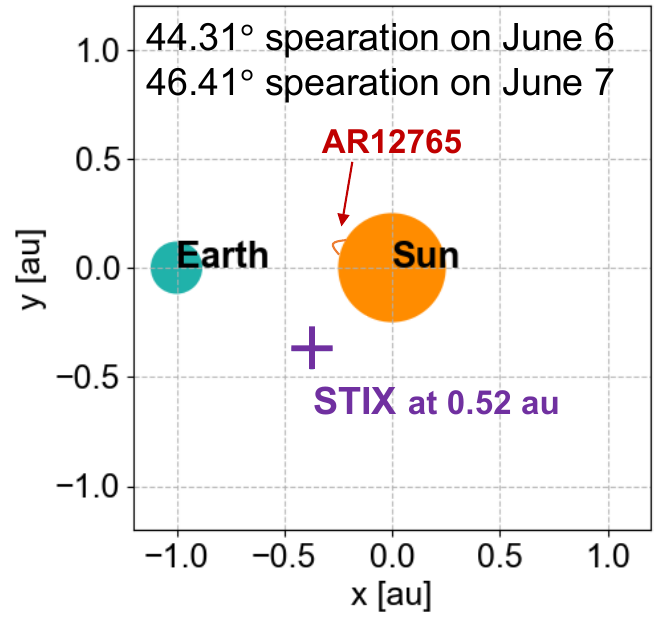}
\caption{The NuSTAR (Earth)--STIX--microflare AR positioning for the observed events. The separation angle defines the approximate NuSTAR--Sun--STIX separation during the two days.}
\label{fig:position_summary}
\end{figure}

\begin{figure*}

  \begin{minipage}[c]{1\textwidth}
  \centering
   \subfloat{%
      \includegraphics[width=0.8\textwidth, trim={0cm 0cm 0cm 0cm},clip]{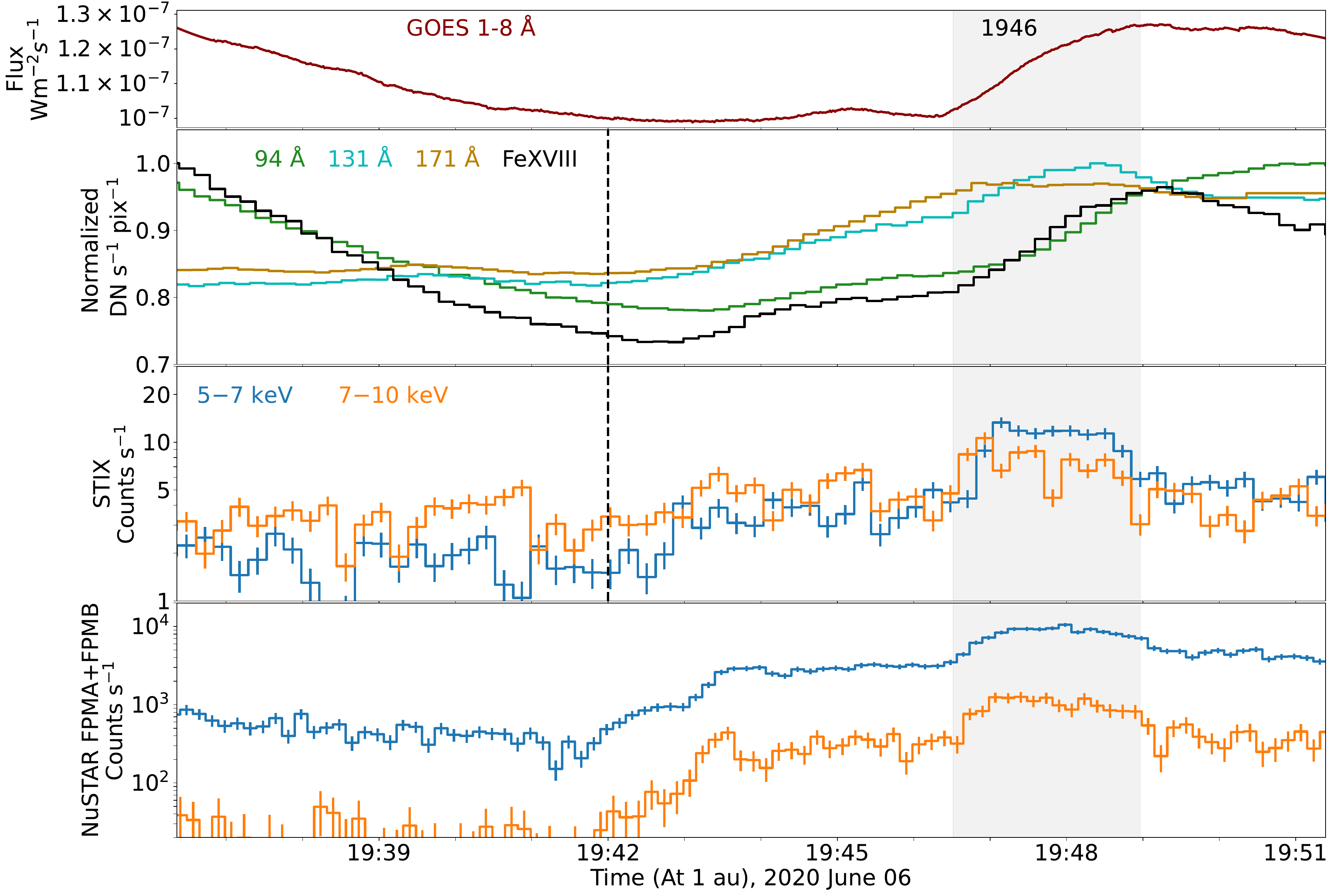}%
    } 

   \subfloat{%
    \includegraphics[width=0.8\textwidth, trim={0cm 0cm 0cm 0cm},clip]{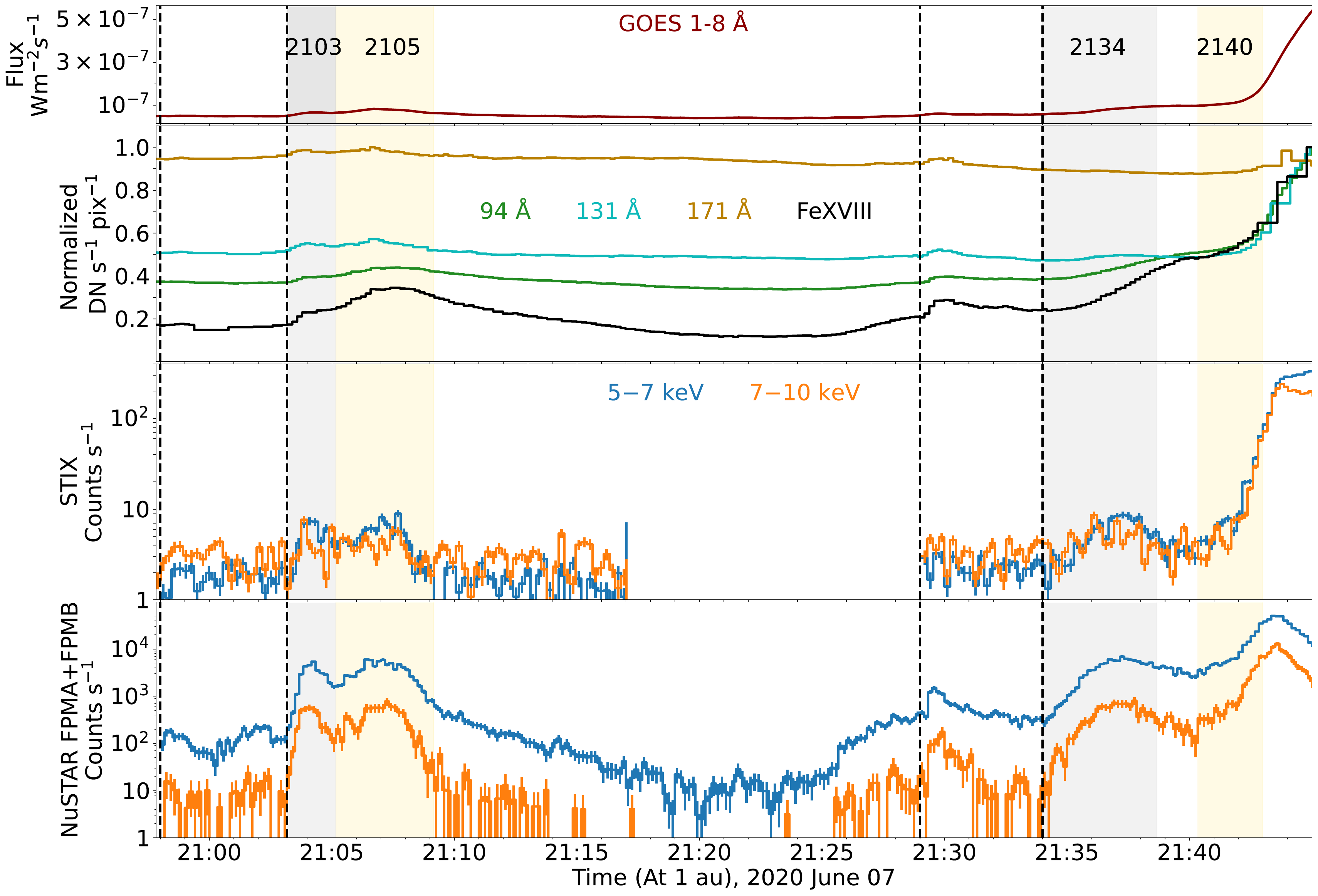}%
  }
  \end{minipage}

\caption{GOES 1--8 {}\AA, SDO/AIA (including Fe XVIII synthetic channel),  STIX (time corrected to 1 au) and NuSTAR (FPMA+B) lightcurves from the 2020 June 6 (top panel) and June 7 (bottom panel) observation campaigns. The event integration times are highlighted in grey and yellow shaded regions. The regions are labelled using HH:MM notation that is used for a reference throughout the analysis. The dotted lines indicate start and end times of pre-flare emission. Due to ghost rays from the decay of a bigger flare close to NuSTAR's FOV, we do not have reliable NuSTAR pre-flare emission for the 1946 event. The STIX pre-flare phase for this event started at 19:34:50, hence only the end time is shown.}
\label{fig:event_summary}
\end{figure*}

On 2020 June 6 and 7, NuSTAR performed four hour-long dwell observations of active region AR12765 located close to the south east solar limb. STIX, during the Solar Orbiter commissioning phase, simultaneously observed the same region, resulting in the first NuSTAR-STIX jointly observed microflares. The positioning of the instruments with respect to the active region is shown in Figure \ref{fig:position_summary}. Despite STIX being in the commission phase, an analysis of 26 early commission phase microflares, done by \citet{Battaglia_2021}, confirms that STIX was working as expected and the collected data can be used for scientific studies. We were able to identify five joint events in two out of the four NuSTAR dwell observations. AR12765 repeatedly flared throughout these two days several occuring outside NuSTAR's observation period, some of which were observed by STIX \citep{Battaglia_2021,Saqri_2022}. 

Figure \ref{fig:event_summary} shows the GOES, SDO/AIA, STIX, and NuSTAR lightcurves. NuSTAR lightcurves contain FPMA+B grade 0 (single pixel) emission in the 5--7 keV and 7--10 keV energy ranges. STIX lightcurves, with same energy ranges as NuSTAR, were time corrected to 1 au. It is clear from the NuSTAR and STIX lightcurves that the observed emission from the two instruments follow very similar evolution which provides a good indication of the consistency between them. The SDO/AIA lightcurves, calculated over the whole AR, show the 94 \AA{}, 171 \AA{} and 131 \AA{} coronal EUV channels as well as Fe XVIII synthetic channel \citep{Del_Zanna_2013}. The 131 \AA{} channel, which has sensitivity to plasma temperatures around 10 MK, best matches the evolution of the HXR emission. This suggests the presence of significant amount of plasma at higher temperatures than majority of previously studied NuSTAR microflares \citep[e.g.,][]{Cooper_2021, Duncan_2021, Wright_2017}. This is expected as the microflares in this paper were of higher GOES A and B class whereas the majority of previously studied NuSTAR microflares were A or sub-A class.

The observed events are some of the brightest microflares that can be observed with NuSTAR as all the event have extremely low, <1\%, livetime (the livetimes are summarised in Table \ref{tab:event_summary}). This results in a variation of NuSTAR's gain \citep{Duncan_2021}, which needs to be corrected for during spectral fitting as described in Section \ref{sec:joint_fitting_06}. This extremely low livetime regime also results in pile-up that needs to be accounted for (see Section \ref{sec:pile-up}). All events analysed in this study required both gain and pile-up corrections.

Unlike in NuSTAR, the observed events were at the lower limit of STIX's detector capabilities. During the analysis, the majority of the spectra were limited to an energy range of 4--11 keV with an exception in the 2140 event that extended up to 15 keV. The limited energy range is partly due to background from the onboard radioactive calibration source (Ba-133). In the low energy range, the emission lines from the source affect the 4--5 keV bin as well as the 7--8 keV bin that is dominated by the escape lines \citep[see bottom panel of fig. 1 in][]{Battaglia_2021}. As a result, the analysed spectra in this study were integrated over longer time intervals in order to improve the counting statistics. Furthermore, the STIX 4--5 keV energy channel is dominated by higher energy photons detected at fractional energies as a result of increasing attenuation from the entrance windows at lower energies \citep[see fig. 1 of][]{Battaglia_2021}. Due to the already high uncertainty associated with this energy channel combined with the high instrumental background, we have decided to only fit the STIX spectra >5 keV. However, it is appropriate to include the 4 keV energy bin for future analysis. The STIX L4 spectrogram files were prepared for spectral fitting using SSWIDL STIX software version 0.5.1. The version 0.5.1 applies the initial version of the STIX transmission calibration. In particular it uses the tungsten grid response derived from optical measurements without taking inflight self-calibration results into account. Furthermore, it uses the nominal value of the thickness of the solar black coating of 5~$\mu m$. The grid transmission used in this analysis is overestimates the photon flux (i.e. the slits are slightly narrower due to stacking errors of the individual layers of the grid), while the solar black thickness underestimates the low-energy photon flux (i.e. the thickness is non uniform therefore effectively less than 5~$\mu m$). Both effects are currently being investigated. A comparison between Fermi/GBM and STIX with the same transmission as used in this paper has shown an agreement between the two instruments within 10\% \citep{Jeffrey_2024}. However, this comparison was done above 20 keV and for the coarse subcollimators only. Therefore, it does not fully translate to the lower energy range in this work.

\section{NuSTAR First Order Pile-up Correction}
    \label{sec:pile-up}
    
    Pile-up occurs if a second photons arrives within the readout time of the first detected photon. In post-processing, this detection is read out as a single photon event with combined energy of the two photons. This can either be caused by a single pixel detecting two photons or a split-pixel event where the second photon is detected in an adjacent pixel of the first one \citep{Grefenstette_2016}. This results in an enhanced count rate at higher energies that needs to be accounted for during spectral fitting. Due to NuSTAR's limited throughput, the incoming photon count rate from the Sun results in detector pile-up which becomes especially problematic for high GOES A and B class flares. 
    
\begin{figure}
    
    	\includegraphics[width=\columnwidth]{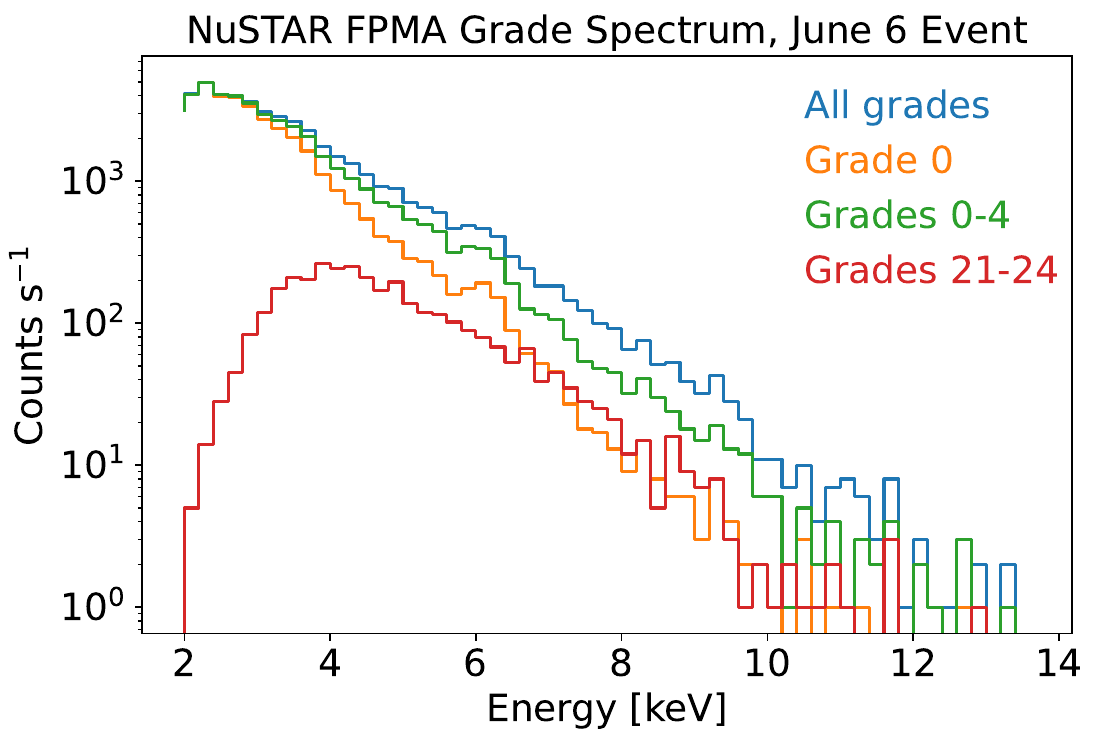}
        \caption{Time integrated NuSTAR FPMA spectrum for the 2020 June 6 event divided into different grades. Grade 0 spectra (orange) are usually used as a full spectrum approximation for solar observation. Due to the extremely low livetime regime this approximation breaks and Grade 0-4 (green) spectrum has to be used instead. Grade 21-24 (red) spectrum is used for modelling pile-up. The all grades spectrum (blue) corresponds to all NuSTAR Grades 0-32.}
        \label{fig:06_nustar_grades}
    \end{figure}

    Each of NuSTAR's detectors consists of a 32×32 pixel array. When an event (photon detection) is triggered on one of the pixels, NuSTAR also records the state of neighbouring pixels \citep{Kitaguchi_2011}. NuSTAR pixel detection is classified into grades. Grade 0 corresponds to single pixel detection. Grades 1--4 are two pixel detections between two horizontally or vertically adjacent neighbouring pixels. These grades are deemed to be either two photon detections or single-photon detection resulting from charge-sharing. Grades 21--24 are two pixel detections with diagonally neighbouring pixels and can only result from a detection of two photons, and hence can be used as an estimate of the pile-up. Different grade spectra for the June 6 microflare are shown in Figure~\ref{fig:06_nustar_grades}, with the substantial Grade 21-24 spectrum clearly indicating pile-up that needs to be removed/corrected.
    
    For sources with non/minimal pile-up, the general analysis approach is to use the Grade 0-4 spectrum for science, with the NuSTAR software (NuSTARADS) providing responses files for these grades. With low pile-up, the Grade 0 spectrum can be used as an approximation of a pile-up free spectrum \citep{Grefenstette_2016}. However for event with significant pile-up a correction instead needs to be applied to the Grade 0-4 spectrum.

    We use the Grade 21--24 spectrum in order to model the pile-up fraction and use it to correct for possible two photon detections contained in Grade 0-4 spectrum. This is done by adding the pile-up model to the other model components (i.e. thermal model) when fitting the Grade 0-4 spectrum. This pile-up model has to be scaled by a factor of 5/4 as we are correcting five grades (Grades 0-4) using a model from a spectrum from four grades (Grades 21-24) \citep[See Appendix C of][]{Grefenstette_2016}. Thus, the pile-up corrected (to first order) Grade 0-4 spectrum only contains single pixel detections and single photon detections resulting in charge sharing.
    
    Empirically, we found that the Grade 21-24 pile-up spectra are best modelled using a photon model given by exponentially modified Gaussian function

    \begin{equation}
    \label{eq:pile-up}
        f(x;\text{a}, \mu ,\sigma ,\lambda )= \text{a} \left({\frac {\lambda }{2}}e^{{\frac {\lambda }{2}}(2\mu +\lambda \sigma ^{2}-2x)}\operatorname {erfc} \left({\frac {\mu +\lambda \sigma ^{2}-x}{{\sqrt {2}}\sigma }}\right) \right)
    \end{equation}

    Where erfc is the complementary error function, and $a$ is a scaling constant, $\mu$ and $\sigma$ are the mean and variance of the Gaussian, and $\lambda$ is the exponential rate. This can also be expressed using a $K$ parameter, which is given by $K=1/\lambda \sigma$ - this form is used by the Python SciPy package\footnote{\href{http://scipy.github.io/devdocs/reference/generated/scipy.stats.exponnorm}{\nolinkurl{http://scipy.github.io/devdocs/reference/generated/scipy.stats.exponnorm}}} which we use for fitting. For this work we use the spectral fitting package as part of the SunPy solar X-ray package, sunkit-spex\footnote{\href{https://github.com/sunpy/sunkit-spex}{\nolinkurl{https://github.com/sunpy/sunkit-spex}}}. Sunkit-spex is currently the only solar-specific software that can perform simultaneous fitting of X-ray spectra from different instruments with the required solar photon models.

    The approach we use for fitting, considering a single NuSTAR FPM for a moment, is to fit the pile-up model given by Equation~\ref{eq:pile-up} to the Grade 21-24 spectrum, whilst simultaneously fitting to the Grade 0-4 spectrum a model comprising of the physical model components (i.e. thermal model) and the pile-up model (scaled by 5/4) with parameters tied to those of Grade 21-24 fit. This can be expanded to fit both FPM spectra, with the simultaneous fit across the four spectra (Grade 0-4 and 21-24 for each of the two FPM), with the physical model parameters tied between the different FPM Grade 0-4 specrta fit (and an additional scaling parameter added to the FPMB model, to represent the relative calibration of FPMA to FPMB).

    Unfortunately there is an additional complication with these low livetime spectra, the need for a gain correction, a slope value (typically a small shift in the range $0.97$ to $1.0$) to divide the energy bins values by, shifting them to their true values \citep{Duncan_2021}. This correction is only possible if there are clear known spectral lines present, and for the microflares studied in this paper this is the case due to the higher temperatures present. So the overall fitting approach we use in this paper is a modified version of that recommended by \citet{Duncan_2021}, to include both pile-up and gain correction. Specifically:
    
    \begin{enumerate}
    \item The first step is to fit the pile-up model to the Grade 21-24 spectra (a separate fit per FPM, but tied to Grade 0-4 spectra of that FPM), whilst simultaneously fitting a thermal model to the Grade 0-4 spectra (the parameters tied across FPM, with scaling factor for FPMB) combined with the pile-up model. In addition a gain correction slope is fitted at the same time across all spectra (with the values tied per FPM). This fit is done over the lower energy range (3-8 keV) where the thermal model dominates, and the pile-up and gain issues are clearly present.
    \item A second step is used if an additional model component is understudy (i.e. non-thermal at higher energies) and/or an additional spectra are available (i.e. STIX). Here again the physical model (i.e. thermal + non-thermal) is fitted simultaneously across multiple spectra (NuSTAR FPMA, FPMB Grade 0-4, and STIX, with scaling factors for FPMB and STIX relative to FPMA), whilst the pile-up model is fitted simultaneously to the NuSTAR Grade 21-24 spectra, and added to the Grade 0--4 model. Here the gain slope correction value found in the first step is used as a fixed value, whereas all the other model parameters (thermal, non-thermal, pile-up) are varied.
    \end{enumerate}

    An example of this approach is shown for the June 6 microflare in Figure \ref{fig:06_joint_fit}, the top row showing the fitted STIX and NuSTAR spectra (Grade 0-4), the bottom row the fitted NuSTAR pile-up spectra (Grade 21-24). These models were fitted using Cash statistics \citep[CSTAT;][]{Cash_1979}{}{} which is commonly used for low count situations, such as we encounter with the NuSTAR spectra. More examples are detailed in sections~\ref{sec:joint_fitting_06} and \ref{sec:joint_fitting_07}.

\section{Joint spectral fitting: 2020 June 6 microflare}
    \label{sec:joint_fitting_06}

    The first detected joint microflare was observed on June 6 between 19:46:31 and 19:48:58 UT, highlighted by the grey region labelled 1946 in the top panel of Figure \ref{fig:event_summary}. It is a GOES B1.2 class flare, with no pre-flare subtraction. The microflare time and livetime are summarised in Table \ref{tab:event_summary}. STIX pre-flare background for this event was integrated between 19:34:50 and 19:42:00 UT. Unfortunately, no pre-flare background was available for NuSTAR due to emission from ghost rays produced by a decaying larger flare from AR12764 that was just outside NuSTAR's FOV. Due the extremely low livetime (<1~s effective exposure), the NuSTAR spectra had to be gain and pile-up corrected. The method used for modelling the pile-up is summarised in Section \ref{sec:pile-up}. 

\begin{figure*}
	\includegraphics[width=0.9\textwidth]{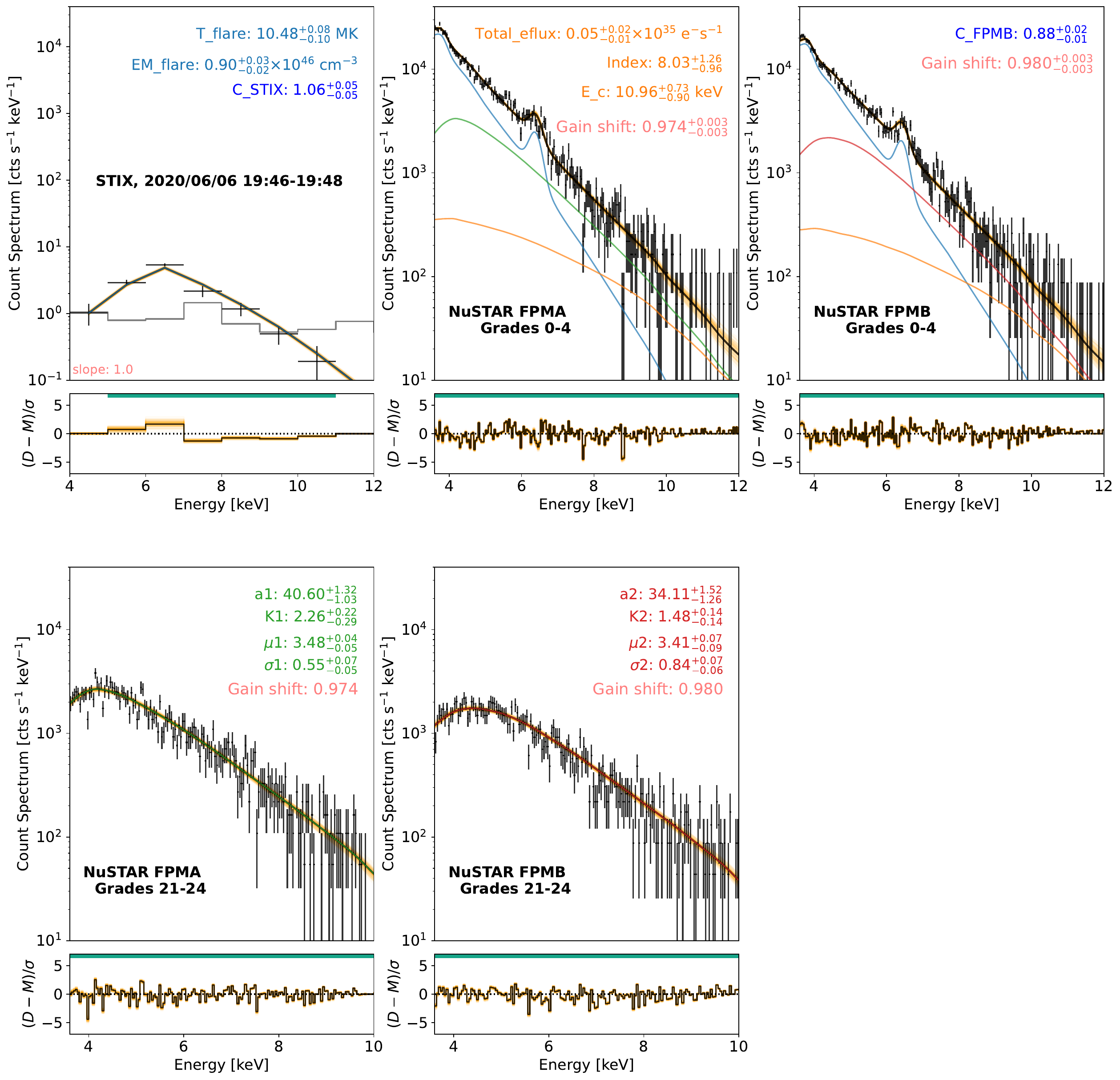}
    \caption{Joint spectral fit of the GOES B1.2 class microflare from 2020 June 6, labelled as 1945 in Figure \ref{fig:event_summary}. STIX (top left) spectrum shows the background subtracted count spectra, with background level shown in grey. The STIX, NuSTAR Grade 0-4 FPMA (top middle) and FPMB (top right) spectra were fitted with a single isothermal model (blue) weighted by a scaling factor C$_{\text{STIX}}$ and C$_{\text{FPMB}}$ with respect to the FPMA spectrum. The NuSTAR spectra we also fitted with an additional non-thermal thick-target model (orange). Furthermore, the NuSTAR Grade 0-4 spectra were pile-up corrected by simultaneously fitting the pile-up model (green and red) to grades 21--24 spectra (bottom panel). All NuSTAR spectra were gain corrected as shown in the plots.}
    \label{fig:06_joint_fit}
    \end{figure*}

    The microflare time integrated spectra for the event (including NuSTAR pile-up fit and gain slope correction) are shown in Figure \ref{fig:06_joint_fit}. The STIX and NuSTAR spectral fitting was performed over energy ranges 5--11 keV and 3.5--12 keV, respectively. The background-subtracted STIX and NuSTAR FPMA\&B spectra were jointly fitted with a single isothermal model (labelled as `f\_vth' in sunkit-spex which consists of thermal bremsstrahlung and line emission, assuming coronal abundances from the CHIANTI atomic database). We fitted the NuSTAR spectra with an additional non-thermal thick-target model (`thick\_fn' in sunkit-spex). Unfortunately, the STIX spectrum is dominated by background in this higher energy range, therefore adding this extra model component is not possible in the joint fit. Therefore, the non-thermal model was only added to NuSTAR. The joint fitting was performed using CSTAT and the 1-$\sigma$ equivalent errors for all the fitted parameters in this study were determined through MCMC analysis available in sunkit-spex.

    The procedure described in section~\ref{sec:pile-up} is used for all the joint spectral fitting in this paper, with the second step modified based on the data. Specifically for the 1946 event, the second step fits the thermal model jointly across NuSTAR Grade 0-4 FPMA, FPMB and STIX (with scaling factors of $C_\mathrm{FPMB}$ and $C_\mathrm{STIX}$ also fitted to determine the relative systematic differences between the spectra), with additional non-thermal component to the NuSTAR Grade 0-4 spectra, whilst simultaneously fitting the pile-up model to the Grade 21-24 spectra, and it being added to Grade 0-4 NuSTAR spectra. The resulting fitted spectra are shown in Figure~\ref{fig:06_joint_fit} and the posterior distribution corner plot for all the fitted parameters is shown in Figure \ref{fig:06_MCMC}.

    The thermal best fit parameters were found to be T~=~$10.48^{+0.08}_{-0.10}$~MK, EM = $0.90^{+0.03}_{-0.02}$ $\times 10^{46}$~cm$^{-3}$. This 10~MK emission explains the consistent time profile between STIX/NuSTAR and the 131\AA{} AIA channel, as the latter is also sensitive to material at this temperature. The best-fit value of the NuSTAR FPMB scaling factor was found to be $0.88^{+0.02}_{-0.01}$ which is within the range of systematic differences expected from the NuSTAR absolute flux calibration \citep{Madsen_2022}. The STIX scaling factor was found to be $1.06^{+0.05}_{-0.05}$ which implies that the systematic difference between the STIX and NuSTAR FPMA spectrum were within only 6\%.

    The thick-target model of a power-law distribution of electrons above a low-energy cutoff was best fitted with total electron flux 0.05$^{+0.02}_{-0.01}$ $\times 10^{35}$ e$^{-}$s$^{-1}$, power-law index $8.03^{+1.26}_{-0.96}$ and low-energy cutoff $10.96^{+0.73}_{-0.90}$ keV. This corresponds to a non-thermal power of $0.98^{+0.31}_{-0.18}$ $\times 10^{26}$ erg s$^{-1}$. Some of the non-thermal parameters have higher associated uncertainties which is not surprising as majority of the energy bins relevant to the non-thermal fit are dominated by pile-up. The value of non-thermal power is within the same order of magnitude as non-thermal power obtained from spectral fitting of a GOES B2 class microflare observed by STIX alone $\sim$30 minutes prior, 19:13--19:33 UT, in the same active region \citep{Saqri_2022}.

\section{Joint imaging: 2020 June 6 Microflare}
\label{sec:joint_imaging}
\begin{figure*}
    \begin{minipage}{0.3\textwidth}
       \subfloat{%
          \includegraphics[width=1\textwidth]{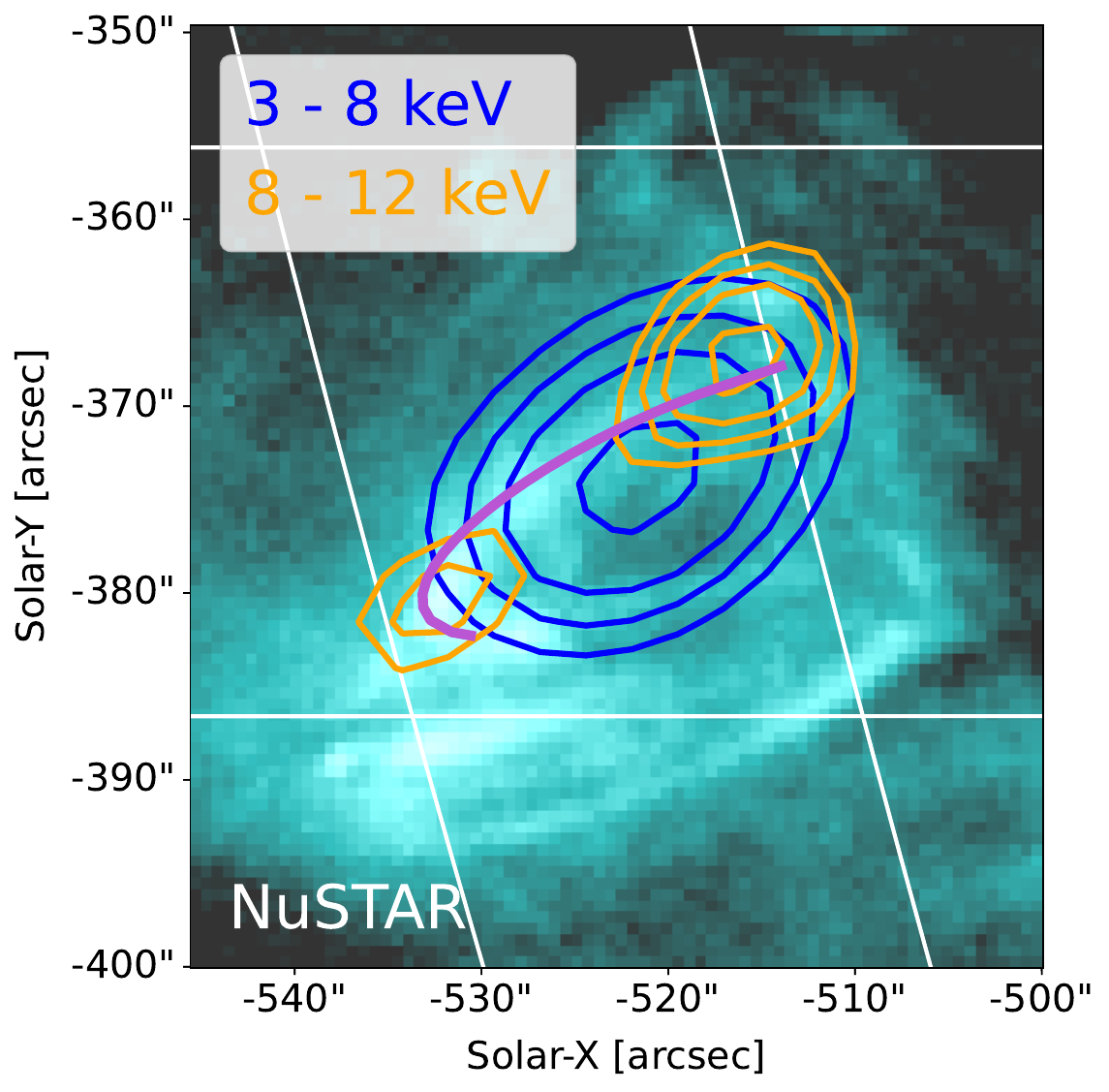}%
          }
    \end{minipage}%
    \begin{minipage}{0.33\textwidth}
        \subfloat{%
           \includegraphics[width=1\textwidth, trim={0cm 0cm 0cm 0cm}, clip]{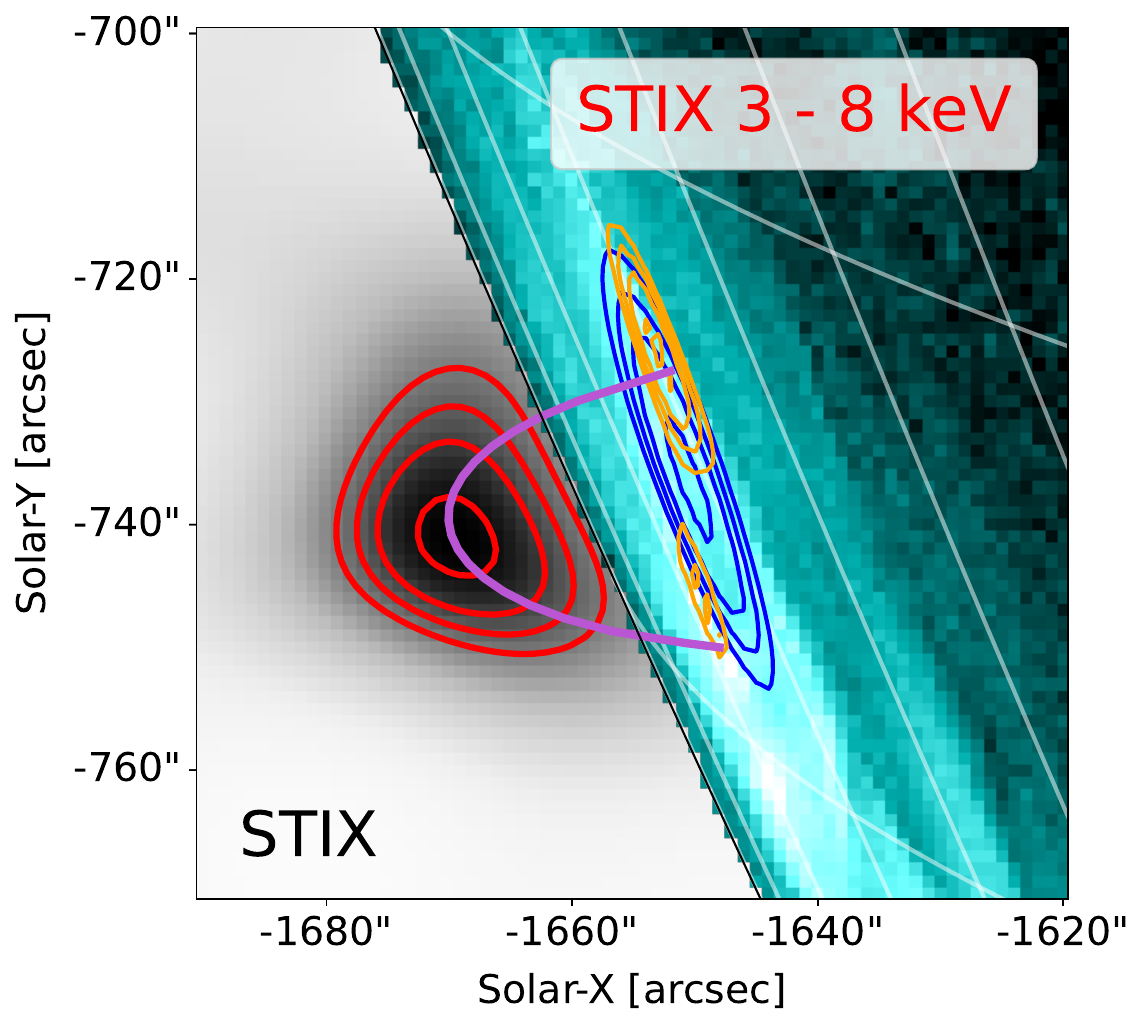}}%
    \end{minipage}
    \caption{Joint imaging for the 1946 microflare. The left panel shows PSF deconvolved NuSTAR FPMA+B image (integrated over the flare time) with 60, 70, 80 and 95\% contour levels overlaid on a flare-time 131 \AA{} SDO/AIA image from 19:47:30 UT. The blue contours are from 3--8 keV (thermal) emission and the orange contours represent the 8--12 keV (non-thermal) emission. A possible flare loop (in purple) was drawn, to highlight the geometry. Right panel shows the STIX image (integrated over 19:46:32--19:47:32~UT) that was reconstructed using MEM\_GE. The red STIX contours (outside the solar limb) have the matching levels to the NuSTAR ones. An 131 \AA{} AIA image, the NuSTAR contours and the flare loop from the Earth view (left panel) were reprojected onto the STIX image (right panel). }
    \label{fig:06_imaging}
\end{figure*}

 Images of NuSTAR and STIX HXR emission are shown overlaid on 131 \AA{} AIA images in Figure \ref{fig:06_imaging}. We performed STIX imaging over the only available 19:46:32--19:47:32~UT time bin and NuSTAR imaging over the entire flare time. The STIX image was reconstructed using a Maximum Entropy Method method \citep[MEM\_GE;][]{Massa_2020}{}{}. The STIX observation only contained $\sim$500 counts which is at the lowest limit of the STIX imaging capabilities, nevertheless we were able to resolve a distinct source. We used the current best aspect solution provided by STIX which is accurate to within 10 arcsec. The source location can be further improved by aligning it with observed AIA emission \citep{Massa_2022}. However, due to the viewing angle, it was hard to estimate any reasonable shift to the STIX source location. The Grade 0 NuSTAR FPMA+B image had the point spread function (PSF) deconvolved (50 iterations using Richardson–Lucy deconvolution) and coaligned with 131 \AA{} AIA emission. 

\begin{figure*}
	\includegraphics[width=0.8\textwidth, trim={0cm 0cm 0cm 0cm}]{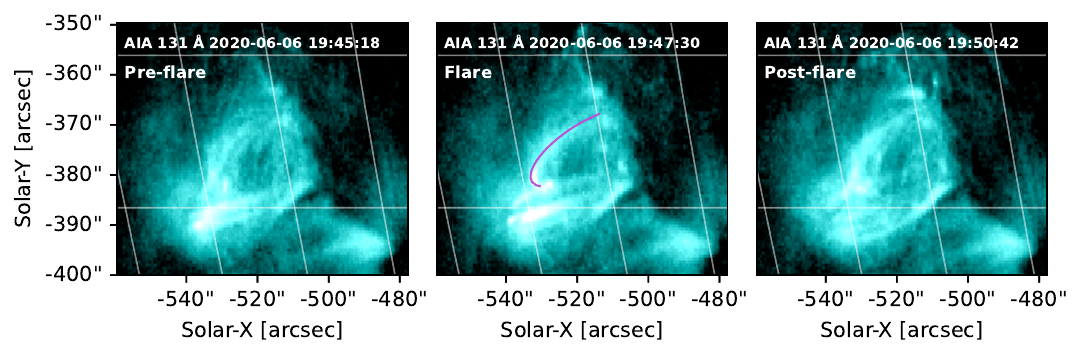}
    \caption{SDO/AIA 131 \AA{} images of the GOES B1 class microflare from 2020 June 6, showing pre-flare (left), flare (middle) and post-flare (right) phases. A possible flare loop (in purple), also shown in Figure \ref{fig:06_imaging}, was added to the flare time panel to highlight the geometry.}
    \label{fig:06_aia_images}
\end{figure*}

The images show 3--8 keV and 8--12 keV emission contours, with the first range being dominated by thermal and the second range being dominated by non-thermal emission (see Figure \ref{fig:06_joint_fit}). The 131~\AA{} AIA image in the left panel of Figure \ref{fig:06_imaging} shows multiple heated loops in different directions throughout the integrated time period. As it can be seen in Figure \ref{fig:06_aia_images}, the bottom loop appears in the 131 \AA{} channel at the onset of the event at 19:44 UT and continues being heated throughout the event. The two top loops, around the NuSTAR contours, become visible in the 131 \AA{} channel at the beginning of the microflare (around 19:46 UT). The configuration of multiple heated loops at different angles throughout the integrated time period made it particularly challenging for co-alignment of the observed NuSTAR emission.

From the NuSTAR contours, we can observe two distinct non-thermal sources (orange) at either ends of an elongated thermal source (blue). This indicates the standard flare configuration consisting of non-thermal footpoints connected by hot flare loops \citep[e.g.,][]{Stiefel_2023, Massa_2022}. For reference, an example loop (in purple), modelled as a semi-circle, was added to the NuSTAR image and was also reprojected onto the STIX image. Despite not applying a shift to the STIX source, the reprojected loop and NuSTAR contours align well with the observed STIX emission. The observed STIX emission appears to originate from the hot coronal loop which is consistent with our joint spectral fitting. 

\section{Joint spectral fitting: 2020 June 07 microflares}
\label{sec:joint_fitting_07}
During the June 7 observation campaign we jointly observed two microflares. The STIX, NuSTAR and AIA lightcurves for this observation are shown in the bottom panel of Figure \ref{fig:event_summary}. Similarly to the 1946 event, all the NuSTAR spectra for these events had to be gain and pile-up corrected due to the very low livetime (pile-up model parameters for all the events are summarised in Table \ref{tab:pile-up}). NuSTAR livetimes for each of the events are summarised in Table \ref{tab:event_summary}.

\subsection{GOES A7.7 microflare}

The first microflare was a GOES A7.7 class microflare (without pre-flare subtraction) at 21:03:10 - 21:09:10~UT. The HXR lightcurves, in the bottom panel of Figure \ref{fig:event_summary}, show a double peaked emission feature observed throughout the microflare. At around 21:03~UT a first compact loop appears in the 131 \AA{} channel (also partly in the 94 \AA{} channel) and continues to increase in brightness as shown in the upper middle panel of Figure \ref{fig:07_small_imaging}. By 21:05~UT this loop cools down corresponding to the first peak of the HXR and 131 \AA{} AIA lightcurves. Subsequently, multiple other loops appear within the same region as shown in the 131 \AA{} image in the upper right panel of Figure \ref{fig:07_small_imaging}. This corresponds to the second HXR lightcurve peak that is mainly prominent in the slightly colder 94 \AA{} and Fe XVIII channels.

Due to the very low STIX count rate, we separated the microflare into two large time ranges and integrated the data over these two ranges to improve the counting statistics. The first time range, labelled as 2103 in Figure \ref{fig:event_summary}, was integrated from 21:03:10 - 21:05:10~UT and the second, labelled as 2105, was integrated over 21:05:10 - 21:09:10~UT. The background time 20:57:00 - 21:03:10~UT was used for STIX pre-flare subtraction and NuSTAR pre-flare model fitting. We fitted the background time integrated NuSTAR spectrum with a single isothermal model with T = 5.69$^{+0.03}_{-0.04}$~MK and EM = 1.10$^{+0.03}_{-0.03}~\times~10^{46}$~cm$^{-3}$ that is used as a fixed model component during NuSTAR spectral fitting (NuSTAR pre-flare model is shown in purple colour in the right panels of Figure \ref{fig:07_small_spectra}). Due to the center of emission intersecting the chip gap of NuSTAR's FPMB module, only FPMA detector was used for spectral analysis.

\subsubsection{2103 event}

We fitted an isothermal model to the joint spectra from the 2103 event shown in the top panel of Figure \ref{fig:07_small_spectra}. The fitting was performed over 5--11 keV and 3--10 keV for STIX and NuSTAR, respectively. The best fit parameters were found to be T = $11.10^{+0.14}_{-0.15}$~MK, EM~=~$0.18^{+0.01}_{-0.01} ~\times~10^{46}$~cm$^{-3}$ and C$_{\text{STIX}}$ = 
 $1.16^{+0.09}_{-0.10}$. The high temperature is consistent with the observations of the 131 \AA{} AIA channel.

 The AIA 131 \AA{} and 1600 \AA{} channels for this event are shown in the middle panels of Figure \ref{fig:07_small_imaging}. The AIA images also include PSF deconvolved FPMA Grade 0 NuSTAR 3--7~kev (blue) and 7--12~keV (orange) contours that were aligned with the 131 \AA{} AIA emission. It is not entirely clear why there is ~7" difference between the contour centroids of the two energy ranges, however we do not expect the shift to be significant as it is below the 18" angular resolution of NuSTAR \citep{Harrison_2013}. The X-ray spectra do not show any clear signs of an additional non-thermal or thermal emission and the 1600 \AA{} AIA channel also does not show any significant brightening in the chromosphere within the region.
 
\subsubsection{2105 event}

As shown in the bottom panel of Figure \ref{fig:07_small_spectra}, the joint spectra of the 2105 event were fitted with a thermal model over 5--11 keV STIX and 3--12 keV NuSTAR energy ranges. Furthermore, the NuSTAR spectrum was fitted with an additional non-thermal thick-target model (similarly to the 1946 event). The resulting parameters were found to be T = $9.38^{+0.08}_{-0.05}$~MK and EM = $0.59^{+0.01}_{-0.02} \times 10^{46}$~cm$^{-3}$ and C$_{\text{STIX}}$ = $1.28^{+0.04}_{-0.06}$. The cooler temperature, compared to the temperature observed during the 2103 event, agrees with the AIA observations that show a more significant increase in the lightcurves of cooler 94 \AA{} and FeXVII channels.  The thick-target model was best fitted with total electron flux 0.06$^{+0.01}_{-0.01}$~$\times~10^{35}$~e$^{-}$s$^{-1}$, power-law index $8.79^{+0.84}_{-1.12}$ and low-energy cutoff $9.58^{+0.55}_{-0.58}$~keV, which corresponds to a non-thermal power of $1.04 ^{+0.19}_{-0.22}$ $\times 10^{26}$ erg s$^{-1}$.

Since this event was integrated over a longer time period, which averages out the time evolution of parameters, we split up the event into two parts and performed NuSTAR only thermal + thick-target model fits in order to get a better understanding of the temporal evolution of the parameters. The first part, 21:05:10--21:07:00~UT, was best fitted with T = $8.30^{+0.41}_{-0.40}$~MK and EM = $0.87^{+0.12}_{-0.18}~\times~10^{46}$~cm$^{-3}$. The thick-target model was best fitted with total electron flux 0.60$^{+0.19}_{-0.16}$~$\times~10^{35}$~e$^{-}$s$^{-1}$, power-law index $9.03^{+0.66}_{-0.81}$ and low-energy cutoff $7.42^{+0.36}_{-0.44}$ keV, which corresponds to a non-thermal power of $8.13 ^{+2.26}_{-1.93}$ $\times 10^{26}$ erg s$^{-1}$. The second part, 21:07:00 - 21:09:10~UT, was best fitted with T = $10.02^{+0.28}_{-0.29}$~MK and EM = $0.41^{+0.04}_{-0.03} \times 10^{46}$~cm$^{-3}$. The thick-target model was best fitted with total electron flux 0.10$^{+0.10}_{-0.04}$ $\times 10^{35}$ e$^{-}$s$^{-1}$, power-law index $9.05^{+1.37}_{-1.65}$ and low-energy cutoff $8.41^{+0.92}_{-0.10}$ keV, which corresponds to a non-thermal power of $1.81 ^{+1.24}_{-0.83}$ $\times 10^{26}$ erg s$^{-1}$. The second time range shows an increase in the temperature/decrease in emission measure as well as significant decrease in the non-thermal power during the decay phase which is expected at this stage of the flare progression \citep[e.g.,][]{Nagasawa_2022}.

\begin{figure*}
\centering

      \includegraphics[width=0.8\textwidth, trim={0cm 0cm 0cm 0cm},clip]{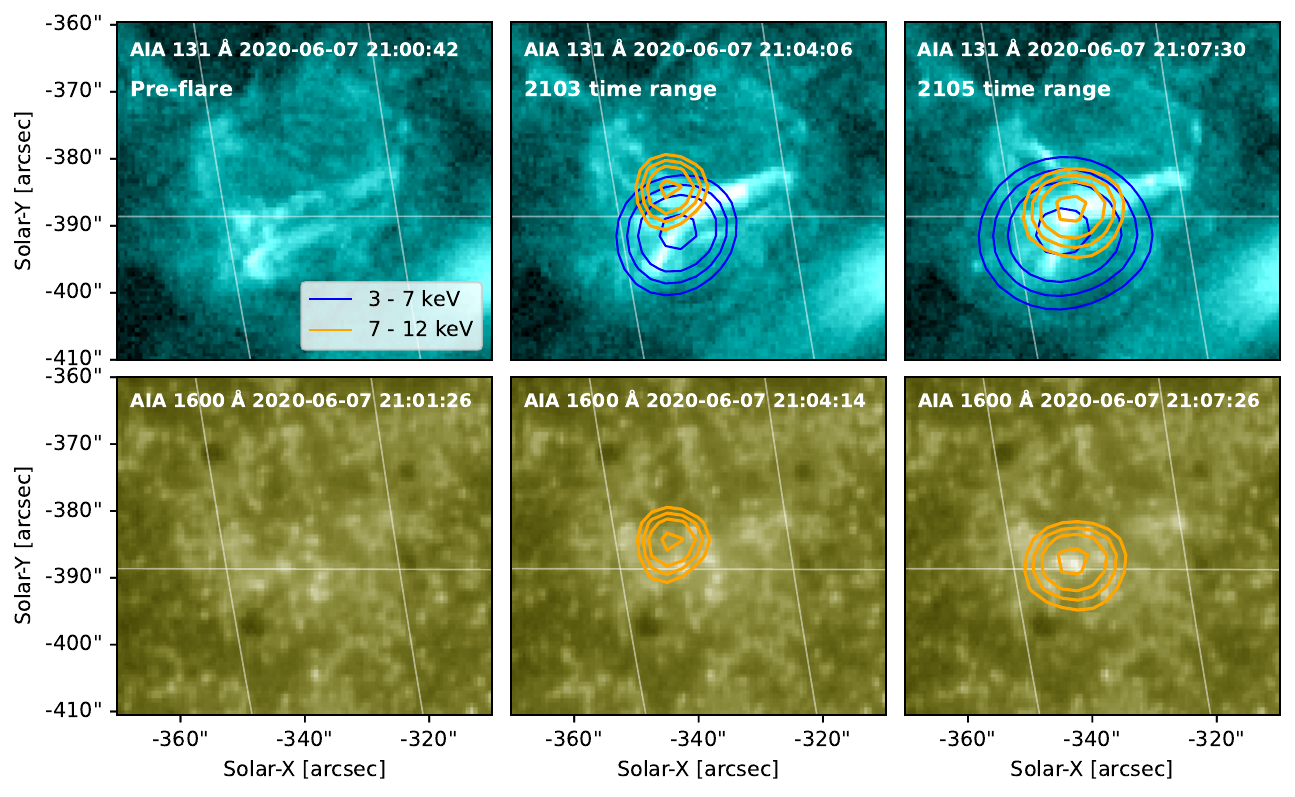}%
\caption{SDO/AIA 131 \AA{} (top panels) and 1600 \AA{} (bottom panels) images of the GOES A7 class microflare from 2020 June 7, showing emission during pre-flare (first column), 2103 (second column) and 2105 (third column) time ranges. The PSF deconvolved FPMA 3--7 keV (blue) and 7--12 keV (orange) emission integrated over the time ranges is overlain on the corresponding AIA panels using  60, 70, 80 and 95\% contour levels. Since the brightenings observed in the 1600 \AA{} panels are very small, we chose to exclude the 3--7 keV contours from those panels to make them more visible.}
\label{fig:07_small_imaging}
\end{figure*}

\begin{figure*}
\centering
\begin{minipage}{1\textwidth}
\centering
        \subfloat{%
           \includegraphics[width=0.9\textwidth]{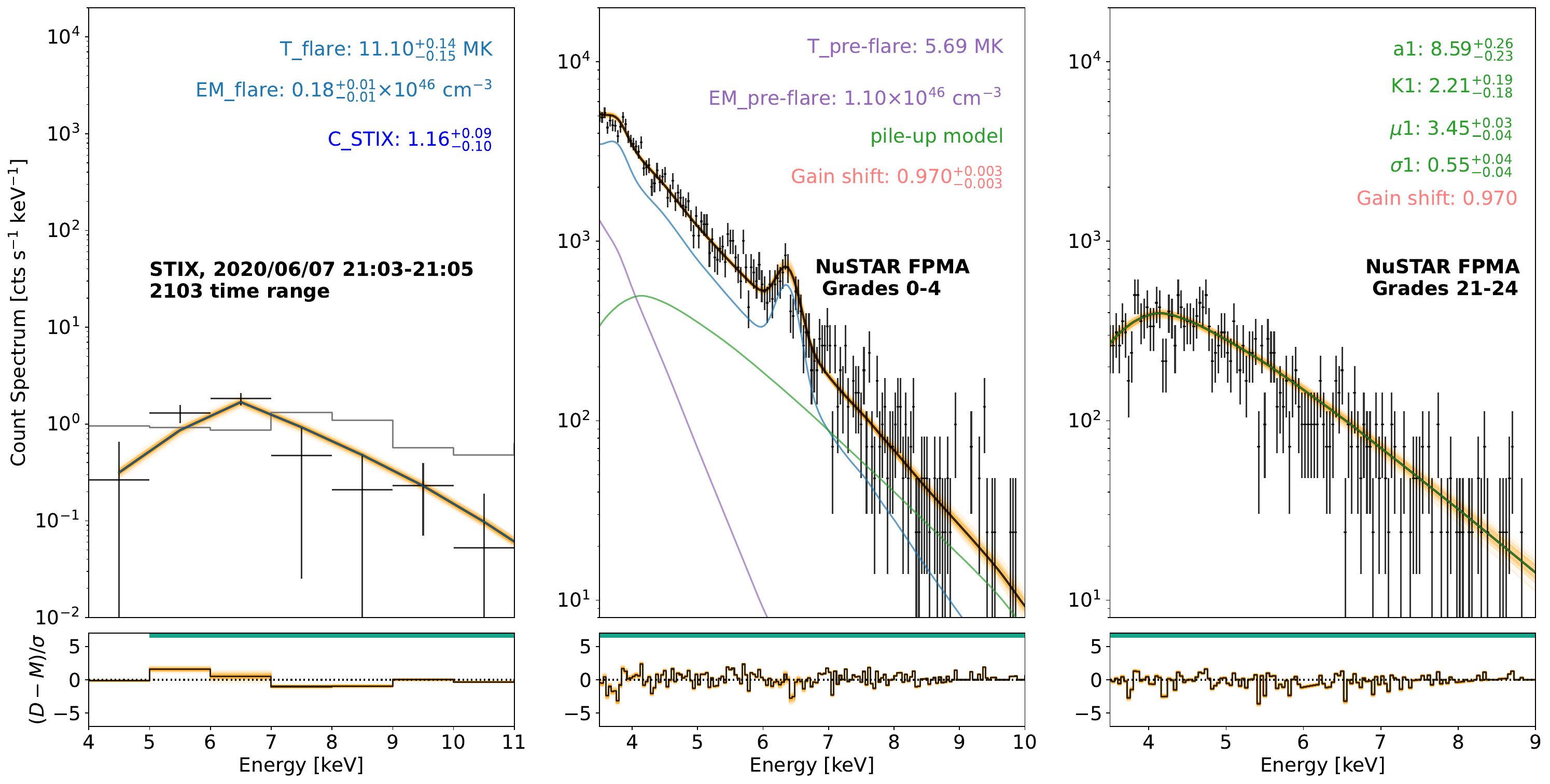}}%

        \subfloat{%
               \includegraphics[width=0.9\textwidth]{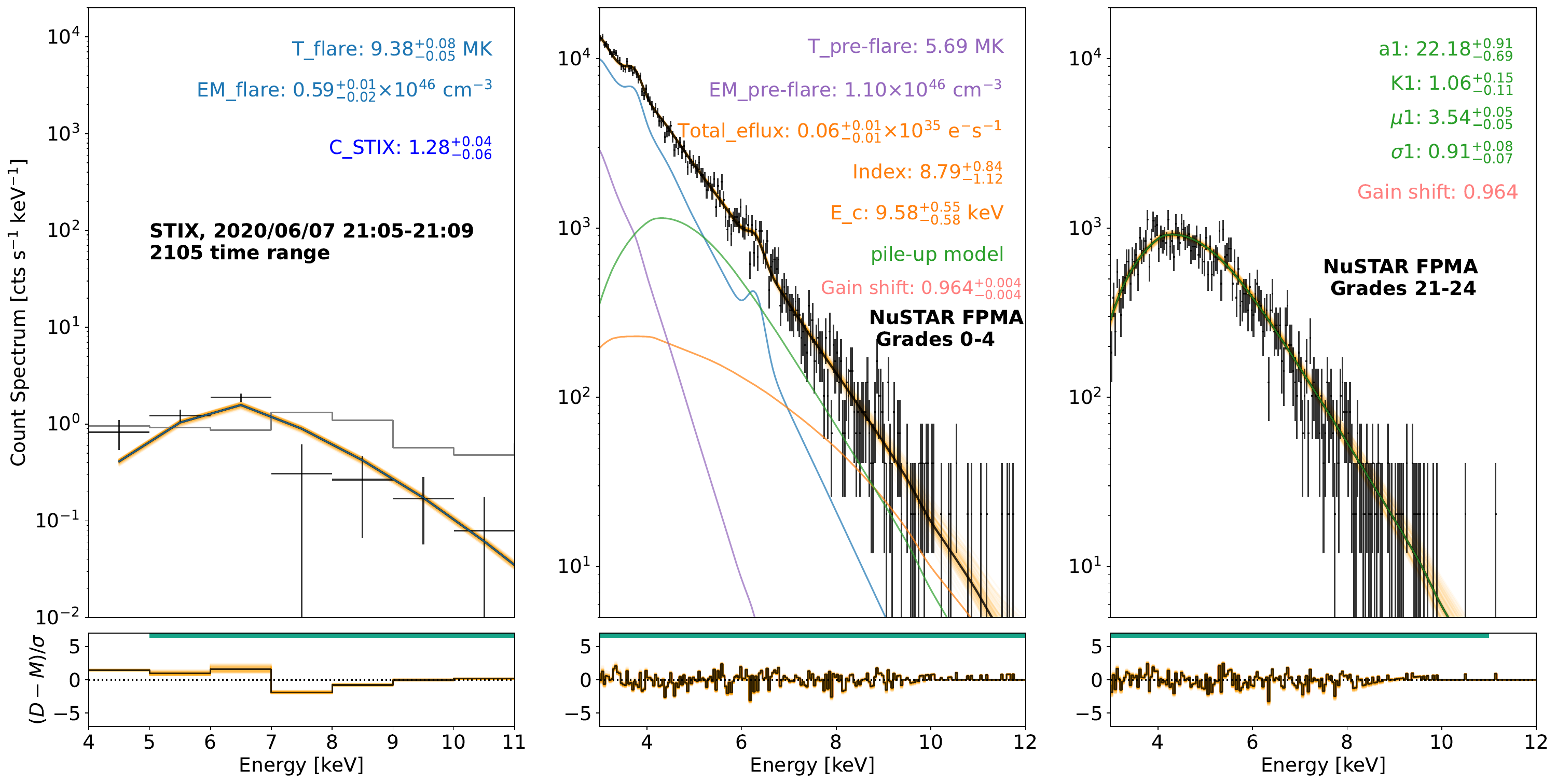}}%
         
    \end{minipage}
\caption{Joint STIX (left panel, background subtracted with background level shown in grey colour) and NuSTAR FPMA Grade 0--4 (middle panel, the isothermal pre-flare model is in purple) and Grade 21--24 (right panel) spectral fits from the 2103 (top) and 2105 (bottom) time ranges. Both time ranges were jointly fitted with a single isothermal model (blue) with STIX scaled by a factor of C\_STIX. The 2105 NuSTAR spectrum was fitted with an additional non-thermal thick-target model (orange). Furthermore, the NuSTAR Grade 0--4 spectra were pile-up corrected by simultaneously fitting the pile-up model (green) to Grade 21--24 spectra. All NuSTAR spectra were gain corrected as shown in the plots.}
\label{fig:07_small_spectra}
\end{figure*}

 The AIA 131 \AA{} and 1600 \AA{} channels for this event are shown in the right panels of Figure \ref{fig:07_small_imaging}. The AIA images also include PSF deconvolved NuSTAR FPMA (Grade 0) 3--7 keV (thermally dominated) and 7--12 keV (non-thermally dominated) contours that were aligned with the 131 \AA{} AIA emission. The position of the thermal emission is almost unchanged compared to the thermal emission from the 2103 event. The non-thermal contours, expected to originate from the chromosphere, align well with the brightening observed in the 1600 \AA{} channel.

\subsection{GOES B6.5 microflare}
The second microflare in this observation window was a GOES B6.4 class microflare \citep[for detailed study of this microflare see][]{Saqri_2022}{}. An extremely  high incident photon rate on NuSTAR detectors after 21:43~UT triggered a photon rejection mode (designed for rejection of cosmic rays). This is the reason for NuSTAR's emission decreasing after 21:43~UT while STIX's emission continues to increase. Therefore only the initial part of impulsive phase could be jointly analysed. Background time 21:29:00--21:34:00 UT was used for pre-flare subtraction for both instruments. The NuSTAR pre-flare isothermal model had best fit parameters T = 5.99$^{+0.08}_{-0.08}$~MK and EM = 1.70$^{+0.10}_{-0.08}\times 10^{46}$ cm$^{-3}$ (NuSTAR pre-flare model is shown in purple colour in Figure \ref{fig:07_big_spectra}). We performed joint fitting of spectra for 2134 (21:34:00--21:38:40 UT) and 2140 (21:40:20--21:43:00~UT) time ranges shown in Figure \ref{fig:event_summary}.

\subsubsection{2134 event}

HXR lightcurves of the 2134 event, shown in Figure \ref{fig:event_summary}, show a sudden increase of the emission that slightly decreases from 21:38--21:40~UT, however the HXR intensity does not return back to the pre-flare level. The 131 \AA{} AIA channel is the only channel that shows a similar feature in its lightcurve. The cooler channels are gradually increasing throughout this phase. 

We fitted the joint spectra with a single isothermal model over 5--11 keV STIX and 3--10 keV NuSTAR energy ranges as shown in the top panel in Figure \ref{fig:07_big_spectra}. The best-fit parameters were found to be T = 11.03$^{+0.08}_{-0.07}$~MK, EM = 0.20$^{+0.01}_{-0.01}$$\times 10^{46}$~cm$^{-3}$ and C$_{\text{STIX}}$ = $1.06^{+0.06}_{-0.05}$.

The AIA 131 \AA{} and 1600 \AA{} channels for this event are shown in the middle panels of Figure \ref{fig:07_big_imaging}. The 131 \AA{} AIA images show a small compact loop that was used for aligning the deconvolved NuSTAR FPMA+B grade 0 3--7 keV (blue) and 7--12 keV (orange) contours. The 1600 \AA{} AIA channel does not show any significant brightening in the chromosphere as it may be expected from the single isothermal model fit.

\subsubsection{2140 event}
\label{sec:07_2140}
The 2140 time range corresponds to the initial part of the impulsive phase of the B6 class microflare. This event was energetic enough for STIX to detect photon energies above the background level up to 15 keV. However, the photon incident rate for NuSTAR was at the limit of the detectors which is shown by the large pile-up, shown in green and red in the bottom panels of Figure \ref{fig:07_big_spectra}, that dominates the spectrum from $\sim$5 keV.

We performed an isothermal + thick-target joint model fit. The spectra and the best-fit models are shown in the bottom panel in Figure \ref{fig:07_big_spectra}. The STIX scaling factor for this fit was found to be C$_{\text{STIX}}$ = $1.52^{+0.09}_{-0.15}$. The isothermal model was best-fitted with T = 10.69$^{+0.12}_{-0.12}$~MK and EM = 0.63$^{+0.03}_{-0.03}\times 10^{46}$~cm$^{-3}$. The thick-target model was best-fitted with total electron flux 0.25$^{+0.06}_{-0.04}$ $\times 10^{35}$~e$^{-}$s$^{-1}$, power-law index $7.97^{+0.07}_{-0.07}$ and low-energy cutoff $7.62^{+0.14}_{-0.13}$~keV. This corresponds to a non-thermal power of $3.53 ^{+0.79}_{-0.56}$ $\times 10^{26}$~erg~s$^{-1}$. As the impulsive phase further progresses and the flare loops continue to be filled with hot plasma, the studied spectra from \cite{Saqri_2022} show an increase of temperature to T=13.4 MK and non-thermal power to $\sim8 \times 10^{26}$ erg s$^{-1}$.

This was a particularly challenging fit as pile-up dominates the NuSTAR spectrum through majority of the energy range which hides the non-thermal model. Thus, the joint fitting with the STIX spectrum provided better constraints on the non-thermal parameters. For simplicity, we neglect directivity effects and albedo in the non-thermal fit as we only fit up to 15 keV. This near-isotropic assumption should still hold within the small energy range (Jeffrey private communication), especially combined with the fact that the background subtracted STIX spectrum is below the 1-$\sigma$ level > 11 keV. Therefore, we would expect a ratio between the NuSTAR and STIX models that is close to 1.

\begin{figure*}
\centering

      \includegraphics[width=1\textwidth, trim={0cm 0cm 0cm 0cm},clip]{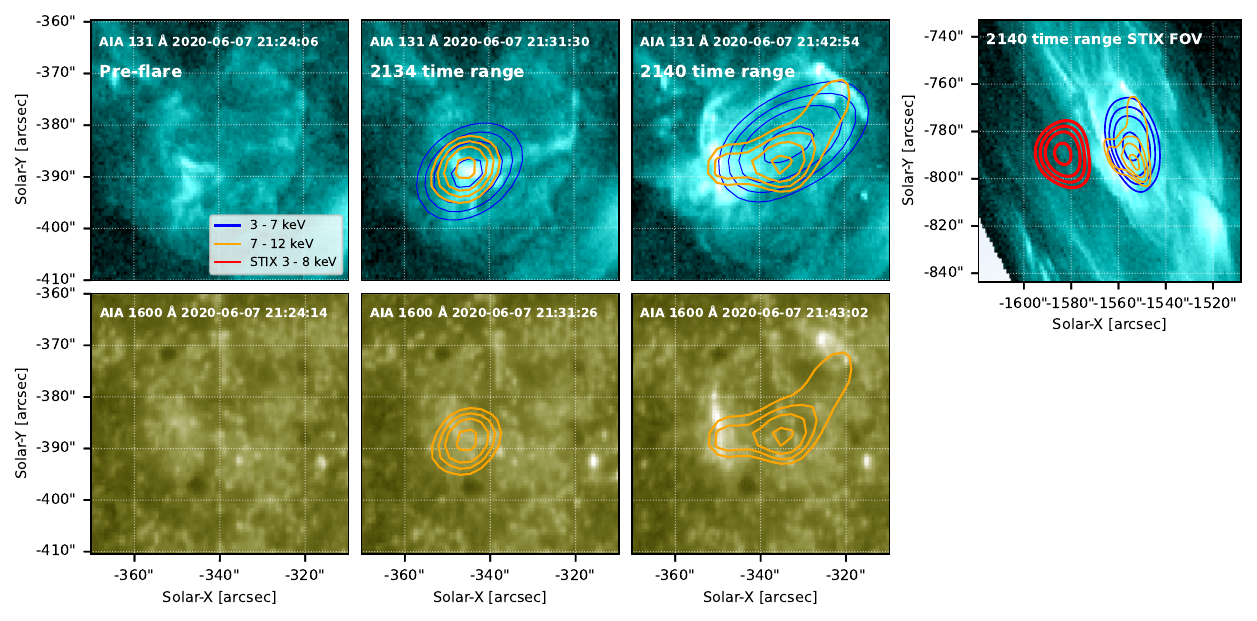}%
\caption{SDO/AIA 131 \AA{}  (top panels) and 1600 \AA{}  (bottom panels) images of the GOES B6 class microflare from 2020 June 7, showing emission during pre-flare (first column), 2134 (second column) and 2140 (third column) time ranges. An additional 131 \AA{} AIA image reprojected onto the STIX FOV is shown in the fourth column. The STIX 3--7 keV, reconstructed using MEME\_GE, and PSF deconvolved FPMA+B 3--7 keV (blue) and 7--12 keV (orange) emission are overlain on the corresponding AIA panels using  60, 70, 80 and 95\% contour levels. STIX imaging was only possible in the 3--8 keV energy range for the 2140 time range. The STIX contours (red) were co-aligned with the NuSTAR 3--7 keV emission. The last panel shows observed STIX emission with reprojected NuSTAR contours. The 3--7 keV contours (in blue for NuSTAR and red for STIX) in 2140 time range panel correspond to thermally dominated emission and the 7--12 keV contours (in orange) correspond to non-thermally dominated emission.}
\label{fig:07_big_imaging}
\end{figure*}

\begin{figure*}
\centering
\begin{minipage}{1\textwidth}
\centering
        \subfloat{%
           \includegraphics[width=0.8\textwidth]{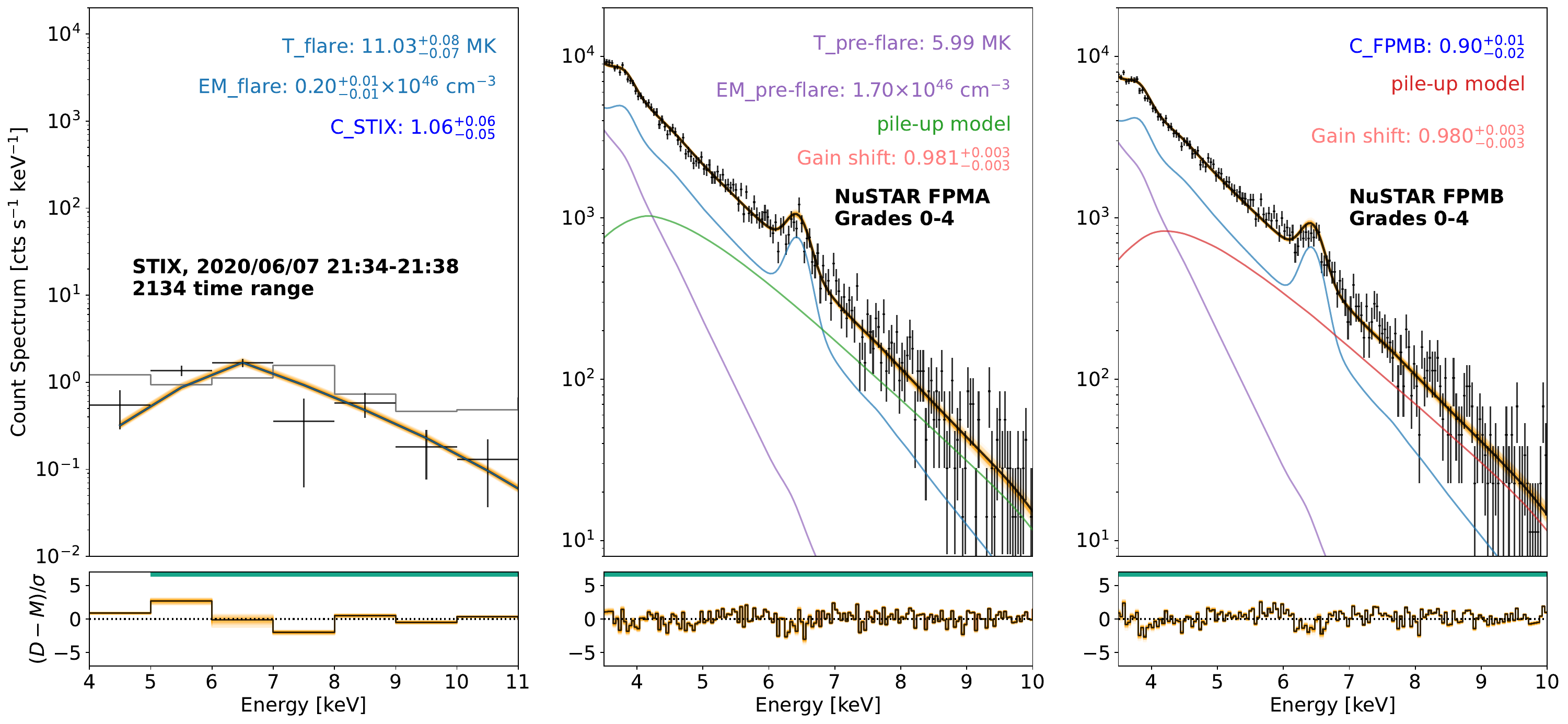}}%

        \subfloat{%
               \includegraphics[width=0.8\textwidth]{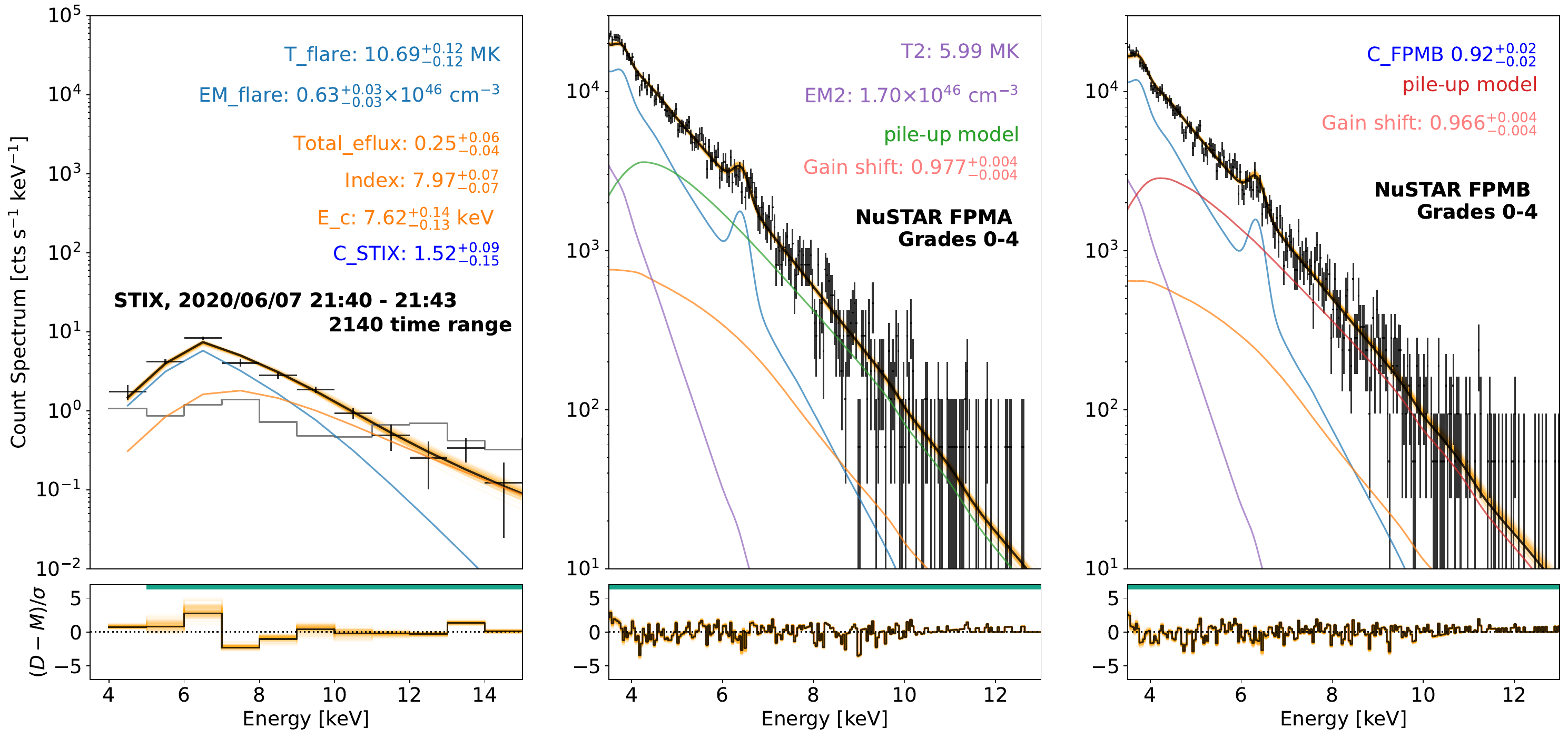}}%
         
    \end{minipage}
\caption{Joint STIX (background subtracted with background level shown in grey colour) and NuSTAR FPMA\&B Grade 0--4 (the pre-flare model is in purple) spectral fits from the 2134 (top panels) and 2140 (bottom panels) time ranges.  The 2134 time range integrated spectra were jointly fitted with a single isothermal model weighted by a scaling factor C$_{\text{STIX}}$ and C$_{\text{FPMB}}$ with respect to FPMA spectrum. The 2140 time range integrated spectra were fitted with a single thermal model (blue) and a non-thermal thick-target model (orange) weighted by the scaling factors. The NuSTAR spectra were gain and pile-up corrected. The pile-up models (green and red) differ for each spectrum (parameters are shown in Table \ref{tab:pile-up}). The pile-up was jointly fitted with Grade 21--24, however, we excluded those spectra from the figure for clarity.}
\label{fig:07_big_spectra}
\end{figure*}

The AIA 131 \AA{} and 1600 \AA{} channels for this event are shown in the right panels of Figure \ref{fig:07_big_imaging}. This was the only event within the June 7 observations with available STIX imaging. Both the deconvolved NuSTAR FPMA+B Grade 0 3--7~keV (thermally dominated) and 7--12~keV (non-thermally dominated) contours are overlain on the AIA image. The NuSTAR contours were aligned using both the hot loops in the 131 \AA{} AIA channel and the brightening at the footpoint region in the 1600 \AA{} AIA channels. The AIA 131 \AA{} image and the NuSTAR contours were then reprojected onto STIX image as shown in the rightmost panel. While the STIX contours, shown in the rightmost panel of Figure \ref{fig:07_big_imaging}, are at a small angle compared to NuSTAR 3--7~keV contours, this is consistent again with a hot loop above the active region. The non-thermal emission has a curved elongated shape which aligns well with the loop footpoints seen in the 1600 \AA{} AIA channel. Unfortunately, the two footpoint sources are not well resolved like for the June 6 microflare, which might be due to the pile-up effects present in the Grade 0 emission which we were not able to correct for during imaging.

\begin{table*}
	\centering
	\caption{Summary of NuSTAR livetime for FPMA (A) and FPMB (B) detectors for all the studied events. The STIX model scaling factor (C$_{\text{STIX}}$) and temperature (T) and emission measure (EM) isothermal model parameters were obtained from the joint fits. Some events were further fitted with an additional non-thermal thick-target model with parameters: electron flux (E\_flux), power law index (Index) and low-energy cutoff (E$_\text{C}$).}
	\label{tab:event_summary}
	\begin{tabular}{ccccccccc} 
		\hline
		 \vtop{\hbox{\strut Event time [UT]}\hbox{\strut}} & \vtop{\hbox{\strut Livetime}\hbox{\strut [\%]}} & \vtop{\hbox{\strut  C$_{\textsc{STIX}}$ }\hbox{\strut}} & \vtop{\hbox{\strut T }\hbox{\strut [MK]}} & \vtop{\hbox{\strut EM }\hbox{\strut [$\times 10^{46} $ cm$^{-3}$]}} & \vtop{\hbox{\strut E\_flux }\hbox{\strut [$\times 10^{35}$ e$^{-}$s$^{-1}$]}} & \vtop{\hbox{\strut Index }\hbox{\strut}} & \vtop{\hbox{\strut E$_\textsc{c}$ }\hbox{\strut [keV]}} & \vtop{\hbox{\strut Non-thermal }\hbox{\strut power}\hbox{\strut [$\times 10^{26}$ erg s$^{-1}$]}}
        \\
		\hline \vspace{1em}
		6-Jun. 19:46:31--19:48:58 $^b$ & \vtop{\hbox{\strut A: 0.31}\hbox{\strut B: 0.39}} & $1.06^{+0.05}_{-0.05}$ & $10.48^{+0.08}_{-0.10}$& $0.90^{+0.03}_{-0.02}$ & $0.05^{+0.02}_{-0.01}$ & $8.03^{+1.26}_{-0.96}$ & $10.96^{+0.73}_{-0.90}$& $0.98 ^{+0.31}_{-0.18}$ \\ \vspace{1em}
  
		7-Jun. 21:03:10--21:05:10 $^a$ &  \vtop{\hbox{\strut A: 0.88}\hbox{\strut}} & $1.16^{+0.09}_{-0.10}$ & $11.10^{+0.14}_{-0.15}$& $0.18^{+0.01}_{-0.01}$ & --- & --- & --- & --- 
        \\ \vspace{1em}
        
        7-Jun. 21:05:10--21:09:10 $^a$$^b$ & \vtop{\hbox{\strut A: 0.69}\hbox{\strut}} & $1.28^{+0.04}_{-0.06}$ & $9.38^{+0.08}_{-0.05}$ & $0.59^{+0.01}_{-0.02}$ &  $0.06^{+0.01}_{-0.01}$ & $8.79^{+0.84}_{-1.12}$ & $9.58^{+0.55}_{-0.58}$ & $1.04 ^{+0.19}_{-0.22}$
        \\ \vspace{1em}
        
        7-Jun. 21:34:00--21:38:40 & \vtop{\hbox{\strut A: 0.63}\hbox{\strut B: 0.78}} & $1.06^{+0.06}_{-0.05}$ & $11.03^{+0.08}_{-0.07}$ & $0.20^{+0.01}_{-0.01}$ & --- & --- & --- & --- 
        \\ \vspace{1em}
        
        7-Jun. 21:40:20--21:43:00 & \vtop{\hbox{\strut A: 0.26}\hbox{\strut B: 0.32}} & $1.52^{+0.09}_{-0.15}$ & $10.69^{+0.12}_{-0.12}$ & $0.63^{+0.03}_{-0.03}$ &  $0.25^{+0.06}_{-0.04}$ & $7.97^{+0.07}_{-0.07}$ & $7.62^{+0.14}_{-0.13}$ & $3.53 ^{+0.79}_{-0.56}$ 
        \\
		\hline
        \multicolumn{7}{l}{$^a$ Only NuSTAR's FPMA was used for spectral fitting.}\\
        \multicolumn{7}{l}{$^b$ The non-thermal fit was only fitted to NuSTAR spectra.}
	\end{tabular}

\end{table*}

\begin{figure}

	\includegraphics[width=\columnwidth]{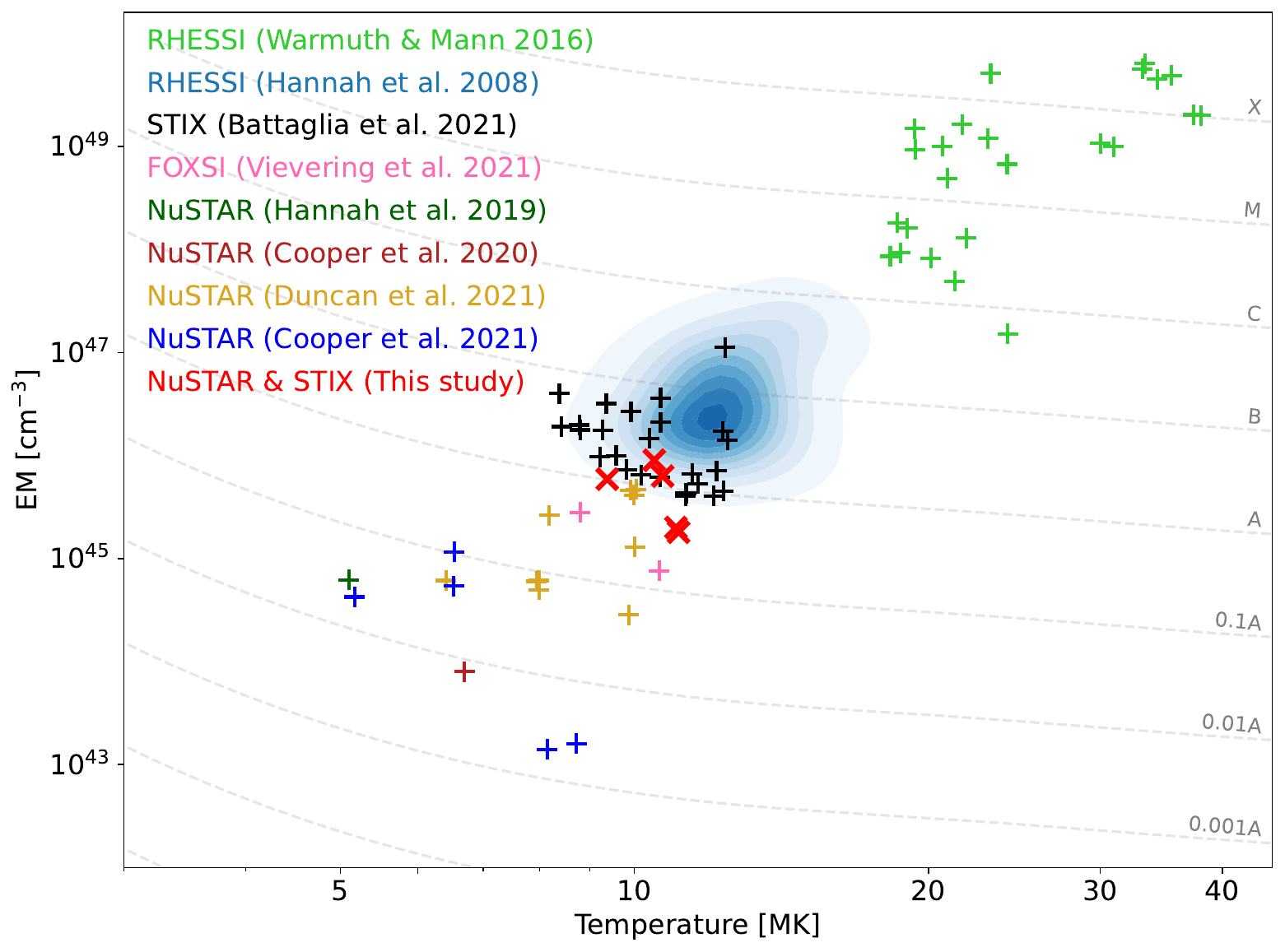}
    \caption{Summary of temperature and emission measure parameters from various NuSTAR \citep[][]{Hannah_2019, Cooper_2020, Duncan_2021, Cooper_2021}{}, STIX \citep{Battaglia_2021}, RHESSI \citep[][]{Warmuth_2016, Hannah_2008}{} and FOXSI \citep{Vievering_2021} flare and microflare studies. The microflares from this study are marked as red crosses in bold.}
    \label{fig:t_em_summary}
\end{figure}

\section{Discussion and conclusion}

In this paper, we present the first joint NuSTAR and STIX spectral and imaging analysis of HXR microflares observed on 2020 June 6 and 7. We performed joint spectral fitting during 5 time ranges across observations of 3 microflares.

The joint spectral fitting was performed over 5--11 keV STIX energy range with an exception to 2140 event that extends up to 15 keV. All STIX spectra were pre-flare background subtracted. The background intervals did not contain any solar signal, but were dominated by counts from the on-board calibration source. A NuSTAR pre-flare model component was only available for the June 7 microflares. The majority of NuSTAR spectral fitting was done between 3--12 keV, with fainter events having an upper energy limit of 10 keV. As part of the NuSTAR analysis, we introduce a new method for NuSTAR pile-up correction that involves modelling pile-up using non-physical 21--24 grades (as shown in the bottom row of Figure \ref{fig:06_joint_fit}). Due to extremely low livetime, all NuSTAR spectra had to be pile-up and gain corrected. This is consistent with analysis of other extremely low livetime NuSTAR events by \cite{Duncan_2021} that have also shown variation in gain. 

The resulting fit parameters from the spectral analysis are summarised in Table \ref{tab:event_summary}. All joint spectra, with an exception to 2140 event, were well fitted with a joint isothermal model with temperatures within 9.38--11.10 MK range. The 2140 was the largest event, impulsive phase of GOES B6 class flare; therefore, we were able to perform joint isothermal + non-thermal thick-target fitting (see Figure \ref{fig:07_big_spectra}). Furthermore, NuSTAR spectra from the 1945 and 2105 events were fitted with an additional thick-target model which was not possible with STIX due to its limited energy range. The non-thermal powers were found to be within the $10^{25}$ -$10^{26}$~erg~s$^{-1}$ range. The values of temperature and emission measure from the analysed events in the context of previous HXR flare studies are shown in Figure \ref{fig:t_em_summary}. As it can be seen from the figure, the isothermal parameters obtained from the joint fits are consistent with other microflare studies and they bridge the gap between the previous STIX and NuSTAR studies. Furthermore, the non-thermal parameters are within the non-thermal constrains of the RHESSI microflares published in \citet{Hannah_2008}. 

The STIX scaling factors for all joint fits were mostly between 6-28\% of NuSTAR's FPMA spectra. This implies that the observed spectra are consistent with each other with only small systematic differences and that both instruments are rather well calibrated. The largest microflare produced a larger scaling factor of 52\%, which indicated the challenges of working with NuSTAR data as the reject rate rises. Overall this is a good result considering that a significant number of the background subtracted STIX count rates were below the background level and the NuSTAR spectra were dominated by pile-up at higher energies. Furthermore, the joint fitting allowed us to leverage different strengths from the two instruments and perform reliable fits for spectra where otherwise we would not be possible (such as when STIX emission was too faint or NuSTAR spectrum was too piled-up). NuSTAR spectra extend to lower energies and have better energy resolution compared to STIX, which is beneficial for fitting thermal models as shown throughout all the events. On the other hand, STIX does not suffer from pile-up like NuSTAR does for these microflares, and therefore we can get better constraints on the non-thermal models that would otherwise be hidden within the NuSTAR pile-up as shown by the 2140 event.

In addition to joint spectral fitting, we were also able to perform joint HXR imaging for the 1946 and 2140 events (as shown in Figures \ref{fig:06_imaging} and \ref{fig:07_big_imaging}). STIX imaging was possible, but limited as expected for these faint microflares during STIX's commissioning phase. In both cases, we were able to resolve one thermal source around the hot flare loops that aligned well with the imaged NuSTAR emission. For the 1946 event, NuSTAR successfully imaged two distinct non-thermal sources (footpoints) on two ends of an elongated thermal source.

There are two main conclusions from this study:
\begin{enumerate}
  \item The model scaling factors implies consistent and well calibrated observations from NuSTAR and STIX. The relatively small scaling factor also implies that the first-order pile-up correction is sufficient at correcting the NuSTAR spectra at the extremely low livetime regime.
  \item All the observed events were at the limits of both telescopes. Higher GOES B class flares observed on-disk are too bright for NuSTAR as the high incoming photon flux triggers rejection mode on-board. Lower GOES B class flares are dominated by high pile-up which is challenging for non-thermal model fitting. On the other hand, lower GOES B and A are very faint for STIX with usable data limited to 4--11 keV and much of the flare spectrum $<1\sigma$. STIX imaging in this low count rate regime is possible but very limited. Thus, the ideal STIX--NuSTAR position configuration for joint observation would be either a fainter GOES A class microflare observed during STIX perihelion or a GOES B class flare which would be observed as occulted for NuSTAR and on-disk for STIX. The second configuration would allow us to possibly detect the faint coronal source (HXR observations closer to the site of energy release and particle acceleration) with NuSTAR whilst also detecting the bright non-thermal footpoints with STIX (which would otherwise be hidden within the NuSTAR pile-up).
\end{enumerate}

Despite the microflare observations being at the limits of both instruments, we successfully performed both joint spectral and imaging analysis of the microflares with consistent results between the two instruments.

\section*{Acknowledgements}

This paper made use of data from the NuSTAR mission, a project led by the California Institute of Technology, managed by the Jet Propulsion Laboratory, funded by the National Aeronautics and Space Administration. Throughout the NuSTAR analysis, we made use of the NuSTAR Data Analysis Software (NUSTARDAS) jointly developed by the ASI Science Data Center (ASDC, Italy), and the California Institute of Technology (USA). Solar Orbiter is a space mission of international collaboration between ESA and NASA, operated by ESA. The STIX instrument is an international collaboration between Switzerland, Poland, France, Czech Republic, Germany, Austria, Ireland, and Italy. SK is supported by the Swiss National Science Foundation Grant 200021L\_189180. The STIX analysis was done using the SolarSoft IDL distribution (SSW) version 0.5.1 from the IDL Astronomy Library. We would like to thank the STIX team for all the valuable insight during the analysis of STIX data. This research also made use of the SunPy open source software package \citep{sunpy_2020} and sunkit-spex, a Python solar X-ray spectral fitting package under the SunPy development.

NB acknowledges support from the UK’s Science and Technology Facilities Council (STFC) doctoral training grant (SST/X508391/1). IGH acknowledges support from a Royal Society University Fellowship (URF/R/180010) and STFC grant (ST/T000422/1). We would like to thank the anonymous referee for their helpful feedback.

\section*{Data Availability}

All data is publicly available. NuSTAR data can be obtained from the NuSTAR Master Catalogue\footnote{\href{https://heasarc.gsfc.nasa.gov/db-perl/W3Browse/w3table.pl?tablehead=name=numaster\&Action=More+Options}{\nolinkurl{https://heasarc.gsfc.nasa.gov/db-perl/W3Browse/w3table.pl?tablehead=name=numaster\&Action=More+Options}}} with OBSIDs 20611002001 and 20611006001. STIX data can be accessed through Solar Orbiter Archive\footnote{\href{https://soar.esac.esa.int/soar/}{\nolinkurl{https://soar.esac.esa.int/soar/}}} with UIDs 1177869888, 1178425664 and 1178416448.



\bibliographystyle{mnras}
\bibliography{example} 




\newpage
\appendix

\section{Pile-up model parameters}

\begin{table}
	\centering
	\caption{Summary of NuSTAR FPMA (A) and FPMB (B) pile-up model parameters given in Equation \ref{eq:pile-up}. The event labels correspond to the time range labels shown in Figure \ref{fig:event_summary}.}
	\label{tab:pile-up}
	\begin{tabular}{ccccc} 
		\hline
		 \vtop{\hbox{\strut Event}\hbox{\strut label}}  & a  & K & $\mu$ & $\sigma$ \\
		\hline
		1946 & A: $40.60^{+1.32}_{-1.03}$ & $2.26^{+0.22}_{-0.29}$ & $3.48^{+0.04}_{-0.05}$ &  $0.55^{+0.07}_{-0.05}$ \\  \\ \vspace{1em}

         & B: $34.11^{+1.52}_{-1.26}$ & $1.48^{+0.14}_{-0.14}$ & $3.41^{+0.07}_{-0.09}$ &  $0.84^{+0.07}_{-0.06}$ \\  \vspace{1em}
  
		2103 &  A: $8.59^{+0.26}_{-0.23}$ & $2.21^{+0.19}_{-0.18}$ & $3.45^{+0.03}_{-0.04}$ &  $0.55^{+0.04}_{-0.04}$ \\  \vspace{1em}
        
        2105 & A: $22.18^{+0.91}_{-0.69}$ & $1.06^{+0.15}_{-0.11}$ & $3.54^{+0.05}_{-0.05}$ &  $0.91^{+0.08}_{-0.07}$ \\ 
        
        2134& A: $16.63^{+1.77}_{-1.13}$ & $1.54^{+0.22}_{-0.22}$ & $3.25^{+0.12}_{-0.18}$ &  $0.73^{+0.11}_{-0.09}$ \\  \\ \vspace{1em}

        & B: $13.54^{+1.17}_{-0.92}$ & $1.74^{+0.36}_{-0.24}$ & $3.38^{+0.11}_{-0.15}$ &  $0.67^{+0.10}_{-0.10}$  
        \\ 
        
        2140 & A: $55.61^{+1.18}_{-1.23}$ & $1.88^{+0.14}_{-0.13}$ & $3.46^{+0.04}_{-0.04}$ &  $0.69^{+0.04}_{-0.04}$   
        \\ \\

        & B: $46.56^{+1.54}_{-1.61}$ & $2.33^{+0.15}_{-0.23}$ & $3.47^{+0.06}_{-0.05}$ &  $0.60^{+0.07}_{-0.03}$  
        \\
		\hline
	\end{tabular}

\end{table}

\section{Posterior distributions from the 1942 event spectral fitting}

\begin{figure*}
\centering

      \includegraphics[width=1\textwidth, trim={0cm 0cm 0cm 0cm},clip]{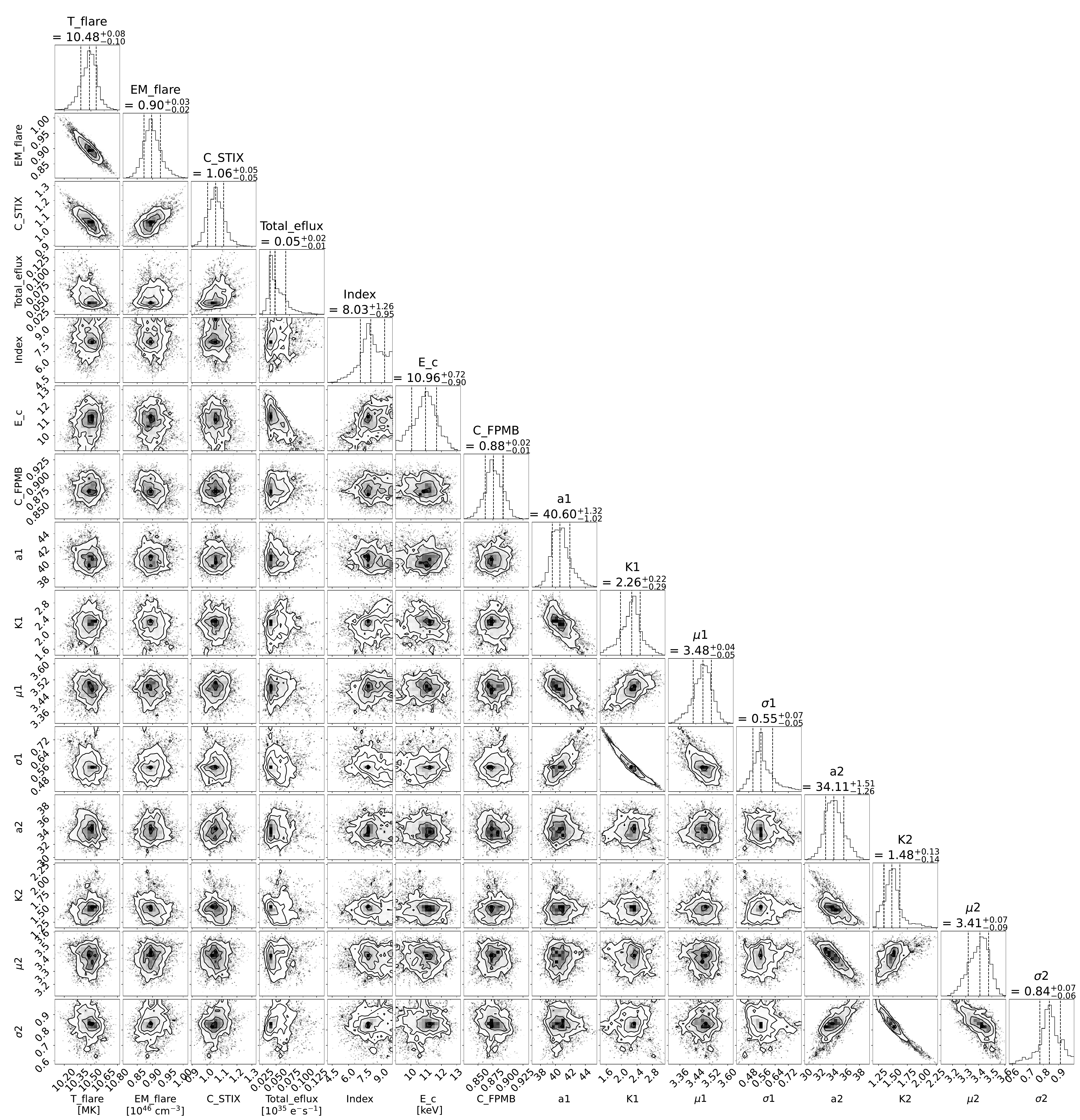}%
\caption{Posterior distributions for thermal, non-thermal thick-target and pile-up models fitted to the 1946 event spectra shown in Figure \ref{fig:06_joint_fit}.}
\label{fig:06_MCMC}
\end{figure*}


\bsp	
\label{lastpage}
\end{document}